\documentclass[useAMS,usenatbib,usegraphicx]{emulateapj}
\usepackage{times}
\usepackage{pifont}
\usepackage{rotating}

\slugcomment{The Astrophysical Journal (Submitted 2015 Feb 02, Accepted 2015 Aug 31).}

\shorttitle{The Fe K$\alpha$ core and dusty torus in AGN}

\def\c{{\em Chandra}}

\def\athena{{\sl Athena}}

\def\astroh{{\sl Astro-H}}

\def\p{$\pm$}
\def\ltsim{\mathrel{\hbox{\rlap{\hbox{\lower4pt\hbox{$\sim$}}}\hbox{$<$}}}}
\def\gtsim{\mathrel{\hbox{\rlap{\hbox{\lower4pt\hbox{$\sim$}}}\hbox{$>$}}}}

\def\Msun{M$_{\odot}$}

\def\aap{A\&A}
\def\nat{Nature}

\def\mnras{MNRAS}
\def\apj{ApJ}
\def\apjs{ApJS}
\def\aj{AJ}
\def\apjl{ApJL}

\def\nh{$N_{\rm H}$}

\def\ha{H$\alpha$}
\def\hb{H$\beta$}

\def\lbol{$L_{\rm Bol}$}

\def\mbh{$M_{\rm BH}$}
\def\gx339{GX~339--4}
\def\swiftj1753{SWIFT~J1753.5--0129}
\def\xtej1118{XTE~J1118+480}

\def\fek{Fe\,K$\alpha$}
\def\rfe{$R_{\rm Fe}$}
\def\rblr{$R_{\rm BLR}$}
\def\rdust{$R_{\rm dust}$}
\def\rgrav{$R_{\rm g}$}
\def\rintf{$R_{\rm dust,intf}$}
\def\rreverb{$R_{\rm dust,rev}$}
\def\taudust{$\tau_{\rm dust}$}
\def\eddratio{$L_{\rm Bol}/L_{\rm Edd}$}
\def\texp{$t_{\rm exp}$}
\def\vfwhm{$v_{\rm FWHM}$}
\def\mytorus{{\sc mytorus}}

\begin{document}

\title{The dust sublimation radius as an outer envelope to the bulk of the narrow Fe K$\alpha$ line emission in Type 1 AGN}
\author{Poshak Gandhi$^1$, Sebastian F. H\"{o}nig$^1$ and Makoto Kishimoto$^2$}
\affil{$^1$School of Physics \& Astronomy, University of Southampton, Highfield, Southampton, SO17 1BJ}
\affil{$^2$Kyoto Sangyo University, Motoyama, Kamigamo, Kita-ku, Kyoto, 603-8555 Japan}

\label{firstpage}
\begin{abstract}
The \fek\ emission line is the most ubiquitous feature in the X-ray spectra of active galactic nuclei (AGN), but the origin of its narrow core remains uncertain. Here, we investigate the connection between the sizes of the \fek\ core emission regions and the measured sizes of the dusty tori in 13 local Type 1 AGN. The observed \fek\ emission radii (\rfe) are determined from spectrally resolved line widths in X-ray grating spectra, and the dust sublimation radii (\rdust) are measured either from optical/near-infrared reverberation time lags or from resolved near-infrared interferometric data. This direct comparison shows, on an object-by-object basis, that the dust sublimation radius forms an outer envelope to the bulk of the \fek\ emission. \rfe\ matches \rdust\ well in the AGN with the best constrained line widths currently. In a significant fraction of objects without a clear narrow line core, \rfe\ is similar to, or smaller than the radius of the optical broad line region. These facts place important constraints on the torus geometries for our sample. Extended tori in which the solid angle of fluorescing gas peaks at well beyond the dust sublimation radius can be ruled out. We also test for luminosity scalings of \rfe, finding that Eddington ratio is not a prime driver in determining the line location in our sample. We discuss in detail potential caveats due to data analysis and instrumental limitations, simplistic line modeling, uncertain black hole masses, as well as sample selection, showing that none of these is likely to bias our core result. The calorimeter on board \astroh\ will soon vastly increase the parameter space over which line measurements can be made, overcoming many of these limitations.
\end{abstract}

\section{Introduction}

The \fek\ line is the most prominent emission feature observed in the spectra of accreting systems. It arises as a result of fluorescence in predominantly cold gas, and has been observed in X-ray binaries as well as active galactic nuclei (AGN). The broad component has been modeled as originating in the inner accretion disk as a result of relativistic broadening \citep{fabian89}. Despite many years of study, though, the origin of the {\em narrow} core of the line in AGN is still unclear. Previous works have found no strong correlation of the Fe line width with those seen in the broad line region (BLR), and suggest that there are likely multiple sites of origin, including the dusty torus, the BLR, as well as the outer accretion disk \citep{yaqoob04, nandra06, shu10}. 

All modern AGN X-ray radiative transfer models include fluorescence emission computed self-consistently for a variety of geometries and obscuring column densities in the accretion disk as well as the torus \citep[e.g. ][]{georgefabian91, pexmon, mytorus, brightmannandra11}, and recent high quality X-ray spectra for many AGN -- obscured as well as unobscured -- bear out the correspondence between transmission, Compton scattering and fluorescence \citep[e.g. ][]{yaqoob12, brenneman14, arevalo14, g14, bauer15, g15_n4785}. Luminosity and covering factor scalings between the infrared and X-ray regimes may also support a close connection between the line-emitting gas and distribution of warm dust \citep{g09_mirxray, ricci14a, toba14}, but any spatial correspondence between the \fek\ emission zone and the torus still remains to be tested in detail. 

The main issue hindering progress in pinning down the origin of the line core is lack of high quality data. In the X-rays, the broad and narrow components of the line must be disentangled, and the narrow core is only resolvable using grating spectra in the brightest AGN at present. In the infrared, the dust emission zones span sub-pc to pc size scales \citep[e.g. ][]{kishimoto11, tristram09, burtscher13}, which require interferometric techniques in order to resolve directly. Infrared reverberation mapping is a growing field and provides an alternate means to infer dust emitter sizes \citep[e.g. ][]{suganuma06, koshida14}. 

There have been recent advances on all these fronts, with a growing number of sources with commensurate data now available. Here we present a comparison of the directly resolved inner edge of the dusty tori (using either interferometry or reverberation mapping) with the sizes of the narrow \fek\ line core regions for a sample of 13 local AGN. We investigate the detailed connection between the two in individual objects, discuss the limitations in the current data, and make comparisons to other recent works. We also test whether the AGN Eddington accretion rates play a role in determining the \fek\ location. Our study, albeit on a small sample, sets the stage for much larger studies which will be possible once the calorimeter on board \astroh\ \citep{astroh14} begins operation in 2016. 

\section{Sample}

\subsection{X-ray grating spectra}
\label{sec:xraysample}

Our starting sample is the \c\ High Energy Grating (HEG) sample of \citet{shu10}, the largest sample to date of high spectral resolution data covering the $\sim$\,6.4\,keV energy regime. This includes 36 unique galaxies below $z$\,=\,0.3. In 27 of these, a measurement of the velocity full widths at half maximum ($v_{\rm FWHM}$) of the narrow line core was reported by the authors. This includes positive FWHM measurements, as well as unresolved line upper limits. The mean $v_{\rm FWHM}$ for this sample was found to be 2060\,\p\,230\,km\,s$^{-1}$. The HEG spectral resolution is 0.012\,\AA, corresponding to a velocity resolution of $\approx$\,1860\,km\,s$^{-1}$ at 6.4\,keV, and is accounted for in the line modeling. 

We have chosen measurements that are likely to best represent the narrow line core. In this respect, note that several new observations for NGC\,4051 have become available since the work of \citet{shu10}. We use the results discussed by \citet{lobban11} and \citet{shu12}. The former work found that the presence of unresolved line component at 6.4\,keV significantly improved the fit of the line complex in the time-averaged data from 12 observations totaling $\sim$\,300\,ks of exposure. Since this component is the one most likely associated with distant material, we treat the core as being unresolved. This leads to an upper limit on $v_{\rm FWHM}$ and a corresponding lower limit on \rfe. 

IC\,4329A is also treated as upper limit on $v_{\rm FWHM}$. Although a broad component to the line is detected by \citet[][see also \citealt{mckernan04}]{shu10} with $v_{\rm FWHM}$\,=\,18830$_{-9620}^{+18590}$\,km\,s$^{-1}$, the line peak energy in this case is found to be $E_{\rm peak}$\,=\,6.305$_{-0.096}^{+0.139}$\,keV. An alternative fit with an unresolved line by the authors yielded a more plausible value of $E_{\rm peak}$\,=\,6.399$_{-0.005}^{+0.006}$\,keV closer to the expected neutral line energy, and the authors stress that these are more reliable measurements of the true narrow core. 

For measurement of the \fek\ emission radii (\rfe), we assume virial motion for the emitting clouds. Then, 

\begin{equation}
R_{\rm Fe}\,=\frac{GM_{\rm BH}}{v^2}
\end{equation}

\noindent
where \mbh\ is the black hole mass, and $v$\,=\,$\sqrt{3}/2$\,$v_{\rm FWHM}$. The correction factor of $\sqrt{3}/2$ arises under the assumption of an isotropic velocity distribution and accounting for the fact that the line-of-sight velocity dispersion is half of the FWHM \citep[cf. ][]{netzer90_saasfee, peterson04}. Uncertainties related to unknown geometric projection factors are discussed below and in the Appendix. Where possible, \mbh\ values based upon reverberation mapping measurements are used, mostly based upon the \hb\ time lag. This was the case for all sources except IRAS\,13349+2438 which is derived from constraints on modeling its spectral energy distribution. In this case, we assumed a large uncertainty equal to the mass measurement itself in order to account for the lack of reverberation mapping. In addition, for NGC\,4151, we use a recent measurement obtained from dynamical modeling. References are given in Table\,1. Uncertainties on \rfe\ were determined by joint Monte Carlo sampling of \mbh\ and $v$ in log space. 

One systematic uncertainty in converting velocities to sizes comes from the unknown geometric projection, which may result in underestimation of the true space velocities, leading to overestimated sizes. This is a well-known effect in BLR reverberation mapping and single-epoch estimates of \mbh, and is captured in the $f$-factor \citep[e.g.][]{peterson04}. If we posit that conservation of angular momentum of the accreting matter leads to a flattened geometry, then we can expect that the projection effects are very similar for the X-ray emitting region and the BLR. In this case, size measurements of these regions will be affected in a similar way for any individual object, so comparisons between these quantities will not be affected. On the sample level, the distribution of geometric inclinations will lead to a widening of any size-luminosity relation above the intrinsic dispersion by a factor of the order unity. Given that our sample covers four decades in luminosity and two decades in sizes, we can expect that the essence of the relation will be preserved even for unknown projection effects in individual sources. In the Appendix, we will discuss alternate size comparisons using reverberation-independent \mbh\ values -- a test which avoids the above assumption.

\subsection{Infrared and optical data}

Near-IR (NIR) observations probe the emission region of the sublimation zone in the torus at temperatures of $\sim$\,1500\,K. The size of this region can be inferred either from reverberation mapping or directly from interferometry. Although both types of measurements are qualitatively similar, the small quantitative differences between both are probably related to the detailed dust distribution \citep[e.g.][]{kishimoto09,hoenig11_variability,koshida14,hoenig14_4151}. Here we collect and use both types of data from the literature (collectively referred to as \rdust). We note that \rdust\ measurements are available mostly for Type 1 AGN, because in Type 2 AGN the innermost hot dust is not easily visible. 

The dust reverberation mapping radii (\rreverb) are the result of a long-term monitoring campaign with the MAGNUM telescope by \citet{koshida14} and have been inferred from $V$- and $K$-band light curve time lags ($\tau_{\rm dust}$) as \rreverb\,=\,$c\tau_{\rm dust}/(1+z)$ where $c$ denotes the speed of light and the $(1+z)$ factor corrects for cosmological time dilation. The interferometric sizes (\rintf) are based on Keck Interferometer data and represent radii of a thin-ring model \citep{kishimoto09,kishimoto11,kishimoto13}.

Of the 27 sources with $v_{\rm FWHM}$ from the sample of \citet{shu10}, 7 have \rintf\ measurements, and 10 have published \taudust\ values. Our final sample of objects with measured values of \rfe\ as well as \rdust\ comprises 13 unique sources. These are listed in Table~1. 

For these objects, we also compute the radii of their optical BLRs (\rblr) for comparison to \rfe. Measurements of the reverberation time lag of the \hb\ emission line ($\tau_{\rm H\beta}$) for 12 AGN of our sample -- i.e. all except IRAS\,13349+2438 -- have been tabulated in \citet{bentz09}, which we use here. Then, \rblr\,=\,$c\tau_{\rm H\beta}/(1+z)$. 

We follow this by investigating any possible relation between the locations of emitting regions and the continuum emission. As a proxy of the latter, we consider the monochromatic continuum luminosities at 5500\,\AA\ ($L_{5500}$), whose values are gathered from \citet{kishimoto11}, \citet{suganuma06} and \citet{bentz09}. These are based upon fitting of their spectral energy distributions and corrected for starlight contamination. A mean uncertainty of 0.1\,dex is assumed for $L_{5500}$. These are also listed in Table\,1.

\section{Results}

\begin{figure*}
  \begin{center}
  \includegraphics[,angle=90,width=16cm]{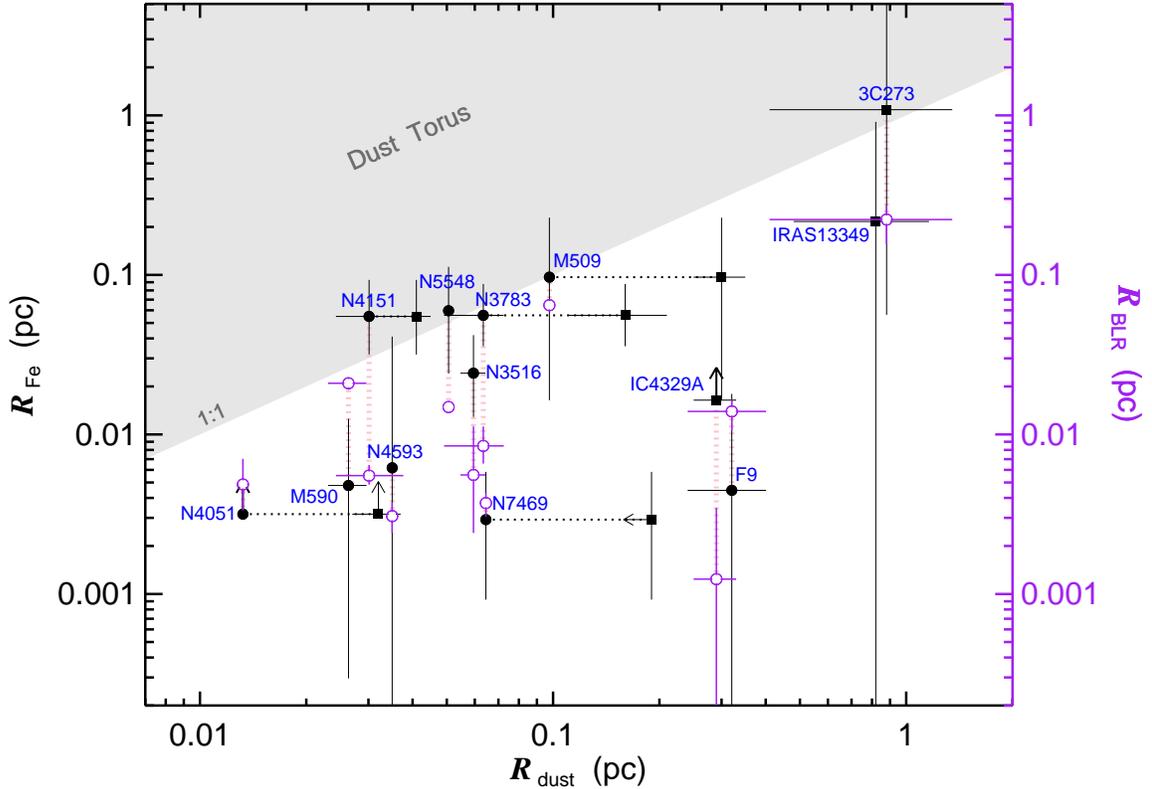}
  \caption{NIR hot dust sublimation radii (\rdust) vs. \fek\ line region radii (\rfe) shown in black. Circles represent \rdust\ from reverberation lags and squares are interferometric measurements. The unfilled violet symbols are the BLR radii (\rblr) based upon optical reverberation mapping. Dotted lines connect these various measurements for any one source. The gray shaded zone represents \rfe\,$>$\,\rdust, i.e. scales commensurate with the body of the torus and beyond. 
    \label{fig:comparison}}
  \end{center}
\end{figure*}

Fig.\,\ref{fig:comparison} shows the comparison between \rfe\ and \rdust. For all objects, the measurements or limits on \rintf\ lie a little above \rreverb. For sources with measurements of both radii, \rintf\ is larger than \rreverb\ by an average factor of 2.3 (\p\,0.3 mean standard deviation). Reverberation is sensitive to the fastest dust response with changing incident radiation, whereas the emission probed in interferometry is more sensitive to the average emitting surface area which is likely to peak at somewhat larger radii \citep[e.g.][]{kishimoto09,hoenig11_variability,koshida14,hoenig14_4151}. In any case, the trends discussed below are similar for both measures of \rdust. 

For 9 of the 13 AGN, the measurements or limits on \rfe\ are fully consistent with \rdust. These are NGC\,3783, NGC\,4151, NGC\,4593, NGC\,5548, Mrk\,509, 3C\,273, IRAS\,13349+2438, NGC\,4051 and IC\,4329A. The remaining 4 sources (NGC\,3516, NGC\,7469, Mrk\,590 and Fairall\,9) have \rfe\ values significantly smaller than \rdust. The ratio of \rdust\,/\,\rfe\ in these 4 ranges over 2.5 (NGC\,3516) to 75 (Fairall\,9). 

The figure also plots \rblr\ on the right-hand axis and reveals some interesting results for individual sources. Whereas the median value of \rblr\ is 8 times smaller than \rfe\ for the full sample, there are 6 objects (NGC\,4051, NGC\,4593, NGC\,7469, Mrk\,509, Fairall\,9 and 3C\,273) for which the \rfe\ measurement or limit is entirely consistent with \rblr. For NGC\,3516, \rfe\ lies below \rdust\ and the uncertainty estimates on \rblr\ and \rfe\ do not overlap, implying an \fek\ origin in an intermediate zone between the BLR and the torus. On the other hand, for Mrk\,590, \rfe\ is significantly smaller than even \rblr. Finally, for Mrk\,509 and Mrk\,590, \rblr\ lies within a factor of 1.5 of \rdust, implying a close proximity of the BLR clouds with the inner extent of the torus. 

Examining the overall distribution of sources, the most important feature is the absence of sources significantly above the line of equality. Whereas there is significant scatter of sources to small values of \rfe, for no source is \rfe\ much greater than \rdust. The two lower limits are also consistent with this line. We discuss these results at length in the next section. 

\begin{figure*}
  \begin{center}
  \includegraphics[angle=90,width=13cm]{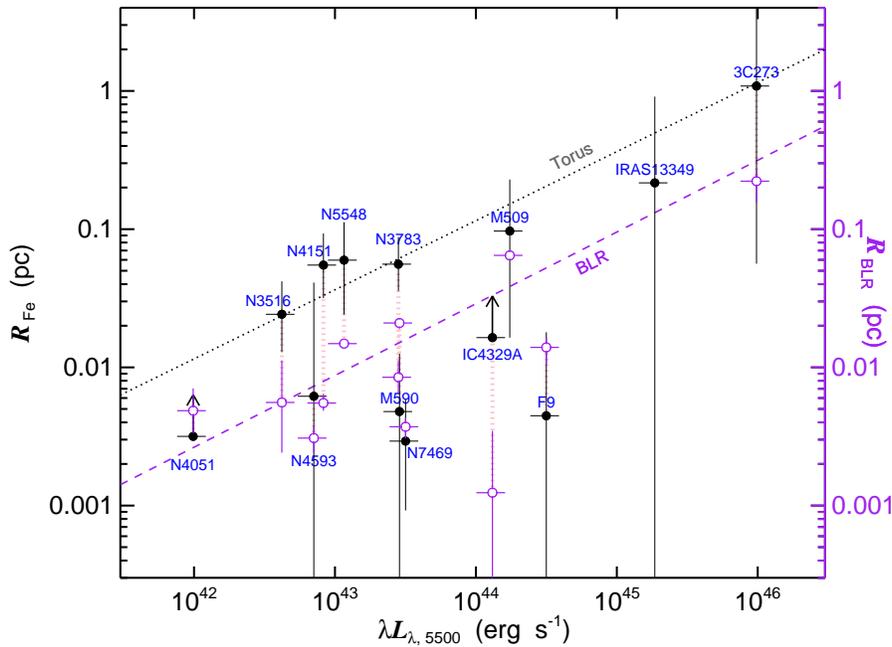}
  \caption{Optical AGN luminosity ($L_{5500}$) vs. \rfe\ (filled) and \rblr\ (unfilled). The dotted line is the torus size--luminosity relation Eq.\,1 from \citet[][originally from \citealt{suganuma06}]{kishimoto11b} and the dashed line is the corresponding relation for the BLR from \citet{bentz09}. 
    \label{fig:fesizes_lopt}}
  \end{center}
\end{figure*}

We next test whether the location of the \fek\ line is driven by fundamental luminosity scalings. We first plot \rfe\ as a function of $L_{5500}$ in Fig.\,\ref{fig:fesizes_lopt}. This shows a close correspondence with Fig.\,\ref{fig:comparison} in all aspects, with an absence of sources on size scales associated with those above the torus size--luminosity relation \citep{barvainis87, kishimoto11}, and a strong scatter of sources below. We tested for any relation between the \rfe\ and $L_{5500}$ with Monte Carlo resampling and computing the distribution of Spearman rank coefficients ($\rho$) for the randomized ensembles. We find $\bar{\rho}$\,=\,0.45$^{+0.21}_{-0.25}$ (90\,\%) and insignificant p-values over the entire ensemble, suggestive of only a weak positive correlation. Any stronger trend is hidden by the large scatter in the current (small) sample. 

\begin{figure*}
  \begin{center}
  \includegraphics[angle=90,width=12.95cm]{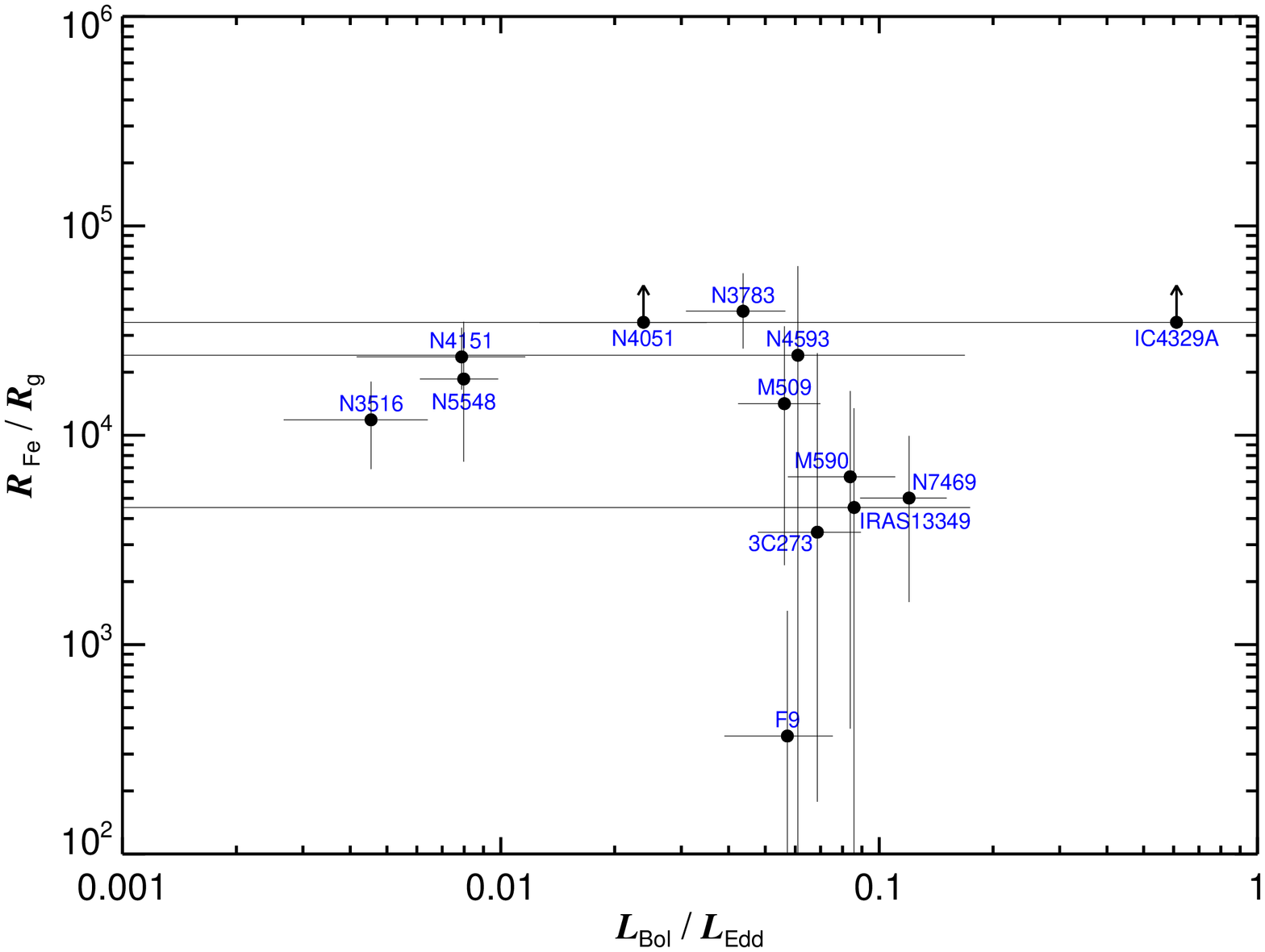}
  \hfill
  \includegraphics[angle=90,width=12.95cm]{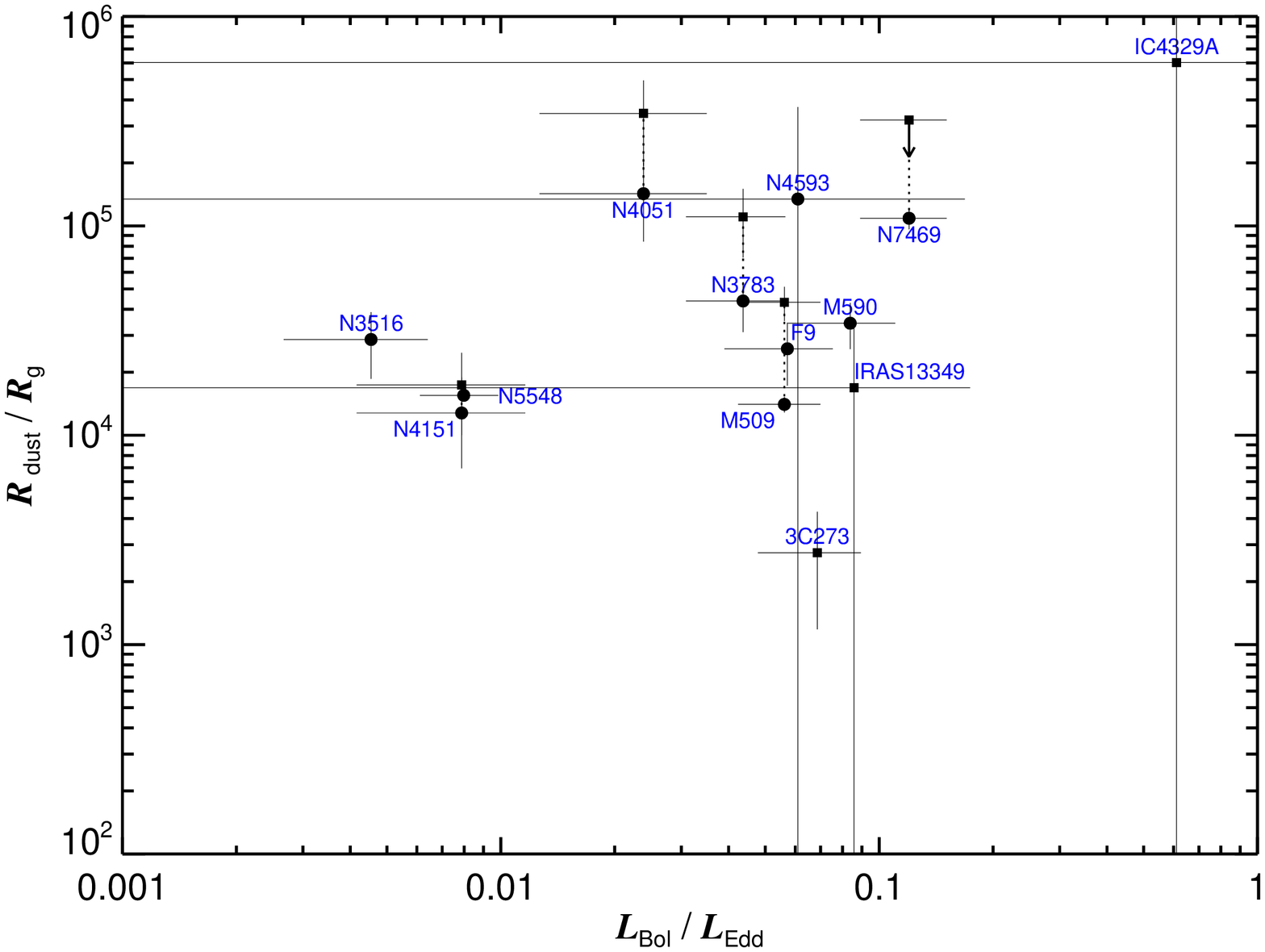}
  \caption{(Top) Eddington ratios (\eddratio) vs. \rfe\ in units of \rgrav. (Bottom) \eddratio\ vs. \rdust\ in units of \rgrav. Symbols are as in Fig.\,\ref{fig:comparison}. 
    \label{fig:rgrav_ledd}}
  \end{center}
\end{figure*}

In Fig.\,\ref{fig:rgrav_ledd}, we remove black hole mass scaling from both axes by plotting \rfe\ in units of the Gravitational radius (\rgrav\,=$G$\mbh/$c^2$) as a function of the Eddington ratio (\eddratio). The latter quantity is taken as a proxy of the (specific) accretion rate. Bolometric luminosities (\lbol) are approximate estimates based upon a correction factor of 6 to $L_{5500}$ \citep{scott04, richards06}. We cross-checked these \lbol\ values with those reported from the broadband modeling carried out by \citet{vasudevanfabian09}, who include all of our sample objects except for IRAS\,13349+2438, and found similar results. Objects with the smallest uncertainties on \rfe\ occupy a narrow region of \rfe/\rgrav\ around $\approx$\,20,000\,km\,s$^{-1}$ (cf. \citealt{shu11}). There appears to be an intriguing hint that sources with high \eddratio\ (above $\approx$\,0.06) show smaller detected values of \rfe/\rgrav\ than sources at low Eddington ratios. But this is not a significant trend in the current sample at least, and for the comparison between \rfe/\rgrav\ and \eddratio\ we find $\bar{\rho}$\,=\,--0.10$^{+0.34}_{-0.42}$ (90\,\%). The second panel in the figure illustrates that dust is present on scales similar to \rfe/\rgrav\ in more than half the sample. 

\section{Discussion}
\label{sec:discussion}

\subsection{Why does \rdust\ serve as an outer envelope to \rfe?}
\label{sec:discussion1}

Our key result is that the dust sublimation radius forms an outer envelope to the \fek\ emission zone in Type 1 AGN, with uncertainties allowing \rfe\ at most a factor of a few times \rdust\ for our sample. This is not an obvious prediction given that the gas distribution is likely to extend continuously on scales both larger and smaller than the innermost dust radius, and that tori may be clumpy with low optical depth \lq holes\rq\ allowing radiation to penetrate well into its body. Our work is thus constraining for AGN torus models, as follows.

For axisymmetric gas distributions, the emitted K$\alpha$ line intensity at any radius $r$ depends upon the local solid angle $\Omega_{\rm K\alpha}(r)$ of gas illuminated by the AGN \citep{krolik87}. The absence of sources with \rfe\,$>>$\,\rdust\ then argues against geometries in which $\Omega_{\rm K\alpha}(r>>R_{\rm dust})$\,$>$\,$\Omega_{\rm K\alpha}(R_{\rm dust})$, i.e. where the line emitting area increases strongly with radius beyond \rdust. Line emission will not cease exactly at \rdust, which would be an unphysical scenario. Fig.\,\ref{fig:comparison} shows that the \rfe\ values of up to a factor of a few times \rdust\ are allowed within the uncertainties. How do typical torus models match up to these constraints? In the geometry adopted by the widely-used \mytorus\ model \citep{mytorus}, the torus is assumed to have a donut shape with a circular cross-section and a half opening angle of 60$^\circ$. Radiation from the nucleus that directly illuminates the torus would impinge upon the donut surface out to distances of the surface tangent line. In the \mytorus\ geometry, this distance is 1.7\,\rdust\ where \rdust\ is the inner edge of the torus. This is consistent within the constraints from Fig.\,1. On the other hand, tori with strong flaring in height as a function of radius would present large illuminated surface solid angles at large nuclear distances, and would not easily satisfy the above constraints on \rfe\ relative to \rdust.

Radiation would also penetrate below the torus surface, but if the integrated column density through the torus\footnote{It should be noted that this is different from the line-of-sight column density, which would vary with inclination angle.} is Compton-thick as is generally thought to be the case, then absorption and Compton-scattering would quickly deplete line photons in the body of the torus, so $\Omega_{\rm K\alpha}$ refers to the solid angle of the reflecting torus surface down to an effective optical depth $\tau_{\rm K\alpha}$\,$\sim$\,1. A natural prediction of this scenario is that sources in which the obscuring tori are likely to be {\em Compton-thin} (e.g. NGC\,2110, \citealt{marinucci15}; NGC\,7213, \citealt{ursini15}) need not follow the trend of Fig.\,\ref{fig:comparison}, i.e. their \fek\ emission need not be restricted to lie around \rdust.

We also emphasize that our results do not exclude the presence of gas on extended scales altogether. If more distant gas is not excited by AGN radiation, it will not fluoresce and will remain invisible. Furthermore, studies based upon line width measurements will selectively target the strongest emission regions present on the innermost (fastest) scales. More distant (and fainter) emitting components could become apparent if our direct line-of-sight to the inner torus were obscured by material optically-thick to the line photons, as may be expected in Type 2 and Compton-thick AGN (cf. the detection of extended emission in NGC\,4945 by \citealt{marinucci12} and in NGC\,1068 by \citealt{bauer15}). Monte Carlo radiative transfer simulations can place detailed constraints on the radial gas geometry in the torus.

There are several potential complexities related to modeling of the \fek\ line and sample selection which must be investigated to test the robustness of the above constraints. These caveats are discussed at length in Section\,\ref{sec:caveats}.

\subsection{\fek\ emission radii much smaller than the torus}
\label{sec:smallscales}

Conversely, the scatter in \rfe\ to scales much {\em smaller} than \rdust\ argues for \fek\ emission from extended regions. This result is not new \citep[cf. ][]{yaqoob04, nandra06, shu10}, but is now inferred from the direct size comparison of Fig.\,\ref{fig:comparison}. The absence of any obvious luminosity scaling in Fig.\,\ref{fig:fesizes_lopt} supports this scenario, and Fig.\,\ref{fig:rgrav_ledd} shows that any Eddington rate--driven physical components (e.g. accretion disk outflows) do not control the origin of \fek, at least over the range of \eddratio\ that we sample. The outer accretion disk and BLR clouds have been invoked as \fek\ emitters in several objects. For example, rapidly variable narrow \fek\ line from small scales have been found in Mrk\,509 \citep{ponti13}, Mrk\,841 \citep{petrucci02} and NGC\,7314 \citep{yaqoob03}. For NGC\,7213, \citet{bianchi08_n7213} found that resolved \fek\ line width matches the optical \ha\ line width and argued for a BLR origin in this case. \citet{shu10} found that a significant fraction of their sample showed \fek\ line widths consistent with, or broader than, the widths of typical optical and infrared photoionized lines, concluding that the \fek\ emission zone location (relative to that of the BLR) genuinely varies from source to source. A BLR origin is also consistent for several sources in our sample. 

Fig.\,\ref{fig:comparison} shows that in many of the sources with the best constraints on \rfe\ (e.g. NGC\,4151, NGC\,5548, NGC\,3783), \rfe\ is well matched to \rdust. However, this does not imply that low grating data S/N is the cause of mismatching \rfe\ and \rdust\ values in other objects. It is certainly true that the largest \fek\ error bars in Figs.\,\ref{fig:comparison}\,and\,\ref{fig:fesizes_lopt} must be unphysical and are likely to be a result of the line fits modeling the underlying continuum (see, for example, the discussion on Fairall\,9 by \citealt{shu10}). In addition, some of the sources with ill-constrained \rfe\ also have relatively short \c\ exposure times (\texp) -- e.g. both Fairall\,9 and NGC\,4593 were observed for 80\,ks with the HEG as compared to a median \texp\,=\,150\,ks for the sample and a maximum of 890\,ks in the case of NGC\,3783. But as emphasized by \citet{yaqoob04}, the uncertainties are not simply a function of S/N, with the grating data showing that the peak \fek\ energy of 6.4\,keV is not dominated by a narrow core in these sources.\footnote{Several of these including Fairall\,9 and NGC\,4593 belong to the \lq Group\,2\rq\ classification of objects in \citet{yaqoob04} with weaker or absent narrow cores.} Moreover, if we use stricter confidence intervals on \vfwhm\ (e.g. the 90\,\% range from \citealt{shu10}), the sources with \rfe\ significantly less than \rdust\ still remain discrepant in all cases. The large uncertainties then appear to reflect an origin from multiple regions in the AGN environment including (fast) clouds present on many scales. Longer grating observations of the sources with large uncertainties on \vfwhm\ will help to confirm the absence of dominant narrow cores, or to disentangle fainter core components. 

\subsection{Comparison to other recent works}

In two other recent works, \citet{jiang11_fek} and \citet{minezaki15} have investigated the use of the \fek\ line for measurement of black hole masses in AGN. \citet{jiang11_fek} assumed a torus origin for the \fek\ line with radii derived from infrared reverberation and found a consistent trend between the masses predicted assuming isotropic motion of the \fek\ clouds with the masses estimated from optical reverberation mapping. However, a scaling factor was required for the two sets of masses to match each other -- the scaling being equivalent to an average line core \vfwhm\ that is 2.6$_{-0.4}^{+0.9}$\,times broader than predicted from initial assumption of a torus origin. 

\citet{minezaki15} studied a restricted AGN sample with \lq best constraints\rq\ on the FWHM of the \fek\ line, and found that generally the \fek\ emission zone is located between the reverberation radii of the broad \hb\ emission line and the dust torus emission, the latter being estimated statistically (i.e. using a fixed ratio of \rdust\ to \rblr\ based on $K$-band reverberation measurements). The criteria that constitute \lq best constraints\rq\ are not quantified by \citet{minezaki15} but their selection excludes most objects with the broadest \fek\ lines, as well as sources with unresolved \fek\ lines. The former category of sources includes objects in which the \fek\ emission zone is consistent with, or significantly smaller than the BLR (as we discussed in Section\,\ref{sec:smallscales}), while the unresolved sources could potentially show \fek\ emission from outside the torus reverberation radius. \citet{jiang11_fek}, on the other hand, do include two sources with upper limits on \vfwhm,\footnote{They treat Mrk\,509 as an upper limit on \vfwhm. The resultant lower limit on \rfe\ is entirely consistent with our trend between \rdust\ and \rfe.} and note the possibility that the narrow line core originates from radii smaller than the infrared radiation. 

Our main result was derived entirely independent of the above works. While being consistent, a key distinction is that we go beyond average comparisons on a statistical basis. Importantly, from our object-by-object comparison, we make a prediction that \rdust\ forms an approximate outer envelope for normal Type 1 AGN. This prediction can be easily falsified if \rfe\ were observed to be much larger than \rdust\ in future observations (see also discussion in next section). In this sense, it is much more powerful than inferences based upon statistical trends. This holds despite the fact that we have not restricted the base sample of \citet{shu10} according to FWHM constraints, and have included upper limits as well. We have then explored the consequences of such a postulation for the structure of AGN tori (Section 4.1) -- aspects not mentioned in these other works whose focus is \mbh\ measurement. We note that since \citet{minezaki15} do not include sources in which the peak \fek\ energy is not dominated by a narrow core,\footnote{Except for their one \lq\lq outlier\rq\rq\ NGC\,7469.} they do not probe  the smallest scales that we discuss in Section 4.2 above. 

Next, we will address various potential caveats which could affect our main result.

\subsection{Potential caveats}
\label{sec:caveats}

Several potential caveats need to be kept in mind regarding current analyses of \c\ \fek\ line profiles, and these are discussed in detail below. 

\subsubsection{Combining multiple HEG datasets}

For several sources (e.g. NGC\,3783, NGC\,4051) multiple HEG datasets have been combined in order to maximize the signal:noise and place the best constraints on the line width. This methodology ignores any potential change in the line emission zone between observations. From the individual  measurements presented by \citet{shu10}, there appears to be some variation in FWHM(\fek) for 2 of the 5 sources with multiple observations spaced by a few months to years (these 2 sources are NGC\,3516 and NGC\,5548). But this is true only at the 68\,\% confidence level, and the 90\,\% confidence intervals show no such variation. In addition, the mean FWHM(\fek) measured {\em per source} (when combining multiple observations) for the sample of \citet{shu10} is entirely consistent with the mean FWHM(\fek) measured {\em per observation}. These facts suggest that combining multiple observations does not strongly affect our results. However, note that for two sources from our sample (3C\,273 and IRAS\,13349+2438), the \fek\ line is detected in only one observation and it is unknown whether this is due to intrinsic and strong variability of the line.

The effective FWHM in a combined dataset can also be artificially broadened if the centroid energy changes between observations. \citet{shu10} find a sharp peak in the rest-frame centroid energies for their sample at $\sim$\,6.400\,keV and quote mean sample line energies centered on the neutral line to within --12 eV and +3 eV (in comparison to the HEG spectral resolution of $\sim$40\,eV), with the {\em per source} analysis fully consistent with the {\em per observation} one. Only in 2 cases out of 32 (3C\,273 and 4C\,74.26) did the centroid deviate strongly from the expected neutral line energy, although they note that these cases are marginal.

We also note that the above centroid energy range is equivalent to systemic velocity shifts of $\approx$\,140 km\,s$^{-1}$ (blueshift) up to $\approx$\,560\,km\,s$^{-1}$ (redshifts) for the fluorescing material. In contrast, the median escape velocity at \rdust\ for our sources is $\sim$2400\,km\,s$^{-1}$ and the median line width for sources with an FWHM detection is 3400\,km\,s$^{-1}$, arguing against any bias related to systemic outflows or inflows.

\subsubsection{The limited HEG spectral resolution}

What is the impact of the limiting spectral resolution of HEG for our results? The velocity resolution of \vfwhm\,=\,1860\,km\,s$^{-1}$ corresponds to \rfe\,=\,0.08\,pc for \mbh\,=\,5\,$\times$\,10$^7$\,\Msun\ (the median \mbh\ for our sample). Whereas much narrower lines can be deconvolved given enough S/N in the data (e.g. see the case of NGC\,3783 which has the best S/N and a resolved line even at 99\,\% confidence), this may not be possible if the S/N is weaker. So are we biased to objects with broader lines, hence creating an artificial limit to \rfe? We argue that this is not the case, because for any given line flux, it is always easier to detect the line if it is unresolved. If the line is significantly broader than the spectral resolution, its peak decreases and the flux is spread over a range of energy bins, in which case detection naturally becomes more difficult. Despite this fact, most sources in our sample already show {\em resolved} lines with the HEG. And we emphasize that we are not excluding objects with unresolved lines which would show only upper limits to \vfwhm. Together, these argue against an artificial limit being created by the HEG spectral resolution in our analysis. 

\subsubsection{Biases due to multiple unmodeled line components}

Another caveat is related to modeling the 6.4\,keV feature as a single Gaussian and ignoring various intrinsic contributions to the feature such as the Compton shoulder which results from down-scattering of line photons off circumnuclear gas. However, the flux in this component is expected to be only $\sim$10--20\,\% of that in the narrow line core for reflection off Compton-thick material \citep{matt02}, so a single Gaussian fit would only be slightly broader than in the absence of a Compton shoulder. We quantified this effect by simulating HEG Fe complex spectra for a range of column densities upto \nh\,=\,10$^{25}$\,cm$^{-2}$ using the {\sc mytorusl} model \citep{mytorus}, which self-consistently models fluorescence and Compton scattering off a toriodal reflector. Another set of simulations using the {\sc gauss} model alone for a single line were carried out for comparison. The intrinsic \fek\ line width was set to $\sigma$\,=\,1\,eV, well below the HEG resolution. We used the standard first order HEG canned response matrices and effective area files\footnote{{\tt http://cxc.harvard.edu/caldb/prop\_plan/grating}} for the simulations and covered a broad range in S/N similar to those in the real data. All spectra were then fit with single Gaussian lines. We found that the effective Gaussian line widths for the simulated {\sc mytorusl} spectra were at most 18\,\% broader than for the single {\sc gauss} models. Correcting for this would increase \rfe\ by at most 39\,\% or 0.14\,dex -- at most a minor effect as can be seen in Fig.\,\ref{fig:comparison}. An independent constraint comes from the fact that \citet{shu10} found a sharp centroid peak for their sample of close to the expect mean rest-frame line energy at 6.400\,keV. If the Compton shoulder were strongly skewing the line fits, the centroid would be expected to lie at lower energies, but this not the case.

Similarly, blended lines could produce artificial broadening in single Gaussian fits. Such lines may arise from multiple ionization stages of Fe, and we here quantify their influence on the measurement of the effective FWHM, and hence on the derivation of \rfe. Firstly, it is worth emphasizing that the observed line centroid energy distribution peaking at $\approx$\,6.40\,keV for the full HEG sample argues strongly for lowly ionized states (typically below Fe\,{\sc xvii}), as discussed by \citet{shu10}. Upon examining the centroid energies ($E$) reported by \citet{shu10} for our sample of 13 AGN, we find a median $E$\,=\,6.403\,keV (\p\,0.010\,keV mean standard deviation). The upper end of this range ($E$\,=\,6.413\,keV) corresponds closely to an ionization stage of Fe\,{\sc xiv} \citep[cf. ][]{liu10, palmeri03, mendoza04}. A further constraint on the strongest ionization stage comes from the distribution of $E$ values for the resolved lines in the longest HEG exposures. There are 5 objects with exposure time of more than $\approx$\,150\,ks: NGC\,3516, NGC\,3783, NGC\,4151, NGC\,5548 and IRAS\,13349+2438. All except IRAS\,13349+2438\footnote{This source is already noted by \citet{shu10} as one of the few anomalous cases with potential line variability between observations. The data for this source also yield only very weak constraints on the overall line properties including flux and FWHM (see Fig.\,1).} show $E$\,$<$\,6.415\,keV at the conservative 90\,\%\ confidence level, again consistent with ionization stages lower than Fe\,{\sc xiv}--{\sc xv}. Based upon this upper energy range, we assess a worst case scenario where all ionization stages from Fe\,{\sc i} to Fe\,{\sc xviv} contribute maximally to artificially broadening the line. All stages are taken to possess the same component line strength and the same intrinsic FWHM. The centroid energies for these stages range over $E$\,$\approx$\,6.394--6.415\,keV \citep[cf. Fig.\,2 of][]{liu10}, or a full range of $\Delta E$\,=\,21\,eV. This corresponds to a velocity broadening of 980\,km\,s$^{-1}$. This velocity should be subtracted in quadrature from the measured FWHM values when modeling the 6.4\,keV feature as a single Gaussian. The largest effect will be on the source with the smallest FWHM for which the line is resolved, which is NGC\,3783 (FWHM\,=\,1750\,km\,s$^{-1}$; see Table\,1), and in this case, the intrinsic FWHM of the individual ionization stages is only 20\,\%\ smaller, at 1450\,km\,s$^{-1}$. The corresponding \rfe\ increases by 45\,\% and is still fully consistent with \rdust. The effect of line blending for the other sources in our sample is much less.

\subsubsection{Sample representativeness}

The parent sample of \citet{shu10} is certainly not a complete sample in a physical sense that is ideally suited for statistical studies. Despite being the largest sample with high quality \fek\ measurements to date, there is some unavoidable bias of HEG observations towards sources with a previously known presence of \fek\ (either broad or narrow) and to objects that were observed with other aims in mind (see also discussion on this point in \citealt{yaqoob04} and \citealt{shu10}). 

On the other hand, we note that there is no bias in terms of physical source properties or classification. The sample includes optical Type 1, 1.5 and 2 AGN and also some Narrow Line Seyfert 1s, all below $z$\,=\,0.3. However, the sample excludes obscured sources with line-of-sight gas column densities \nh\,$>$\,10$^{23}$\,cm$^{-2}$ in order to avoid spectral modeling complexities. While a signal-to-noise requirement is imposed so that the total counts in the full HEG bandpass must be more than 1500, \citet{shu10} state that relaxing this criterion would have admitted only 2 more objects, so this does not appear to be a major restriction unless \fek\ properties evolve very rapidly with redshift (i.e. to fainter source fluxes) -- this is an unlikely prospect in our opinion. 

Of the 36 sources studied by \citet{shu10}, the statistical quality of the data was good enough for FWHM measurements in 27 objects. In the remaining 9 objects, the signal:noise of the data around the Fe line region is so poor that the line itself is not detected at more than 95\,\% confidence and the Gaussian model fit can become unphysically broad as it begins to model the continuum itself (see Section 3 of \citealt{shu10}). This is the case for NGC\,526a, Mrk\,705, NGC\,3227, Mrk\,766, PDS\,456, IRAS\,18325--5926, NGC\,7314, Ark\,564 and MR\,2251--178. Examining the reason for the poor data quality of these 9 sources, we note that their mean HEG exposure time is only 85\,ks, as opposed to 192\,ks for the 27 objects with FWHM measurements. In addition, the two sources with the longest exposure times (PDS\,456 and MR\,2251--178 with exposures of 145.2\,ks and 148.7\,ks, respectively) are both known to be relatively distant, obscured quasars showing highly complex continua and evidence of significant absorption lines resulting from strong outflows \citep[e.g. ][]{gibson05, hagino15}. In other words, the absence of a significant \fek\ emission line in these 9 sources is most likely a combination of relatively short exposure times resulting in low signal:noise, and complex spectra in which a uniform analysis does not immediately resolve a narrow line core.

The main restriction imposed by the infrared selection is one of target visibility -- most (though not all) of the dust reverberation lags have been carried out from the MAGNUM telescope in Hawaii. There may be an implicit preference for observing bright sources and those sources with known (significant) levels of optical variability, but it is not clear how this would bias our sample. Interferometric size measurements have been carried out mostly from Keck also in Hawaii, but the sample of AGN where \rintf\ has been measured is currently limited. So there is no obvious physical bias in the source selection, albeit our sample of only 13 AGN from the parent sample remains small.

\subsubsection{Alternate black hole mass estimates}

Finally, one may question the robustness and suitability of our adopted \mbh\ measurements. As discussed in Section\,\ref{sec:xraysample}, certain assumptions about the geometric correction ($f$-factor) are implicit in virial \mbh\ estimates based upon reverberation mapping, and we have assumed that similar correction factors apply for the \fek\ emission zone. 

The influence of making such an assumption can be judged by using alternate black hole mass measurements, which are entirely independent of reverberation mapping. In the Appendix, we have compiled such measurements in Table\,\ref{tab:msigma}. We present a figure showing the resultant comparison of \rfe\ with \rdust\ in this case, and demonstrate that our main inference of \rdust\ forming an envelope to \rfe\ remains unchanged. We refer the reader to the Appendix for this test.

~\par
\noindent
To summarize, we have discussed and quantified some caveats to our work, including those from combining multiple datasets, limited instrumental resolution and biases related to simplistic (single Gaussian) modeling of a likely complex line feature. None of these issues appears to cause a bias strong enough to overwhelm our main result of \rdust\ acting as an outer envelope to \rfe. We point out current shortfalls in terms of sample completeness, although the apparent heterogeneous nature of sample selection argues against any strong selection bias. We also explore the use of alternate black hole masses and their influence on our main result, which do not change our main result. 

Although we cannot rule out the presence of a {\em combination} of these various biases, it would be surprising if they were to conspire to produce a limit of \rfe\ around \rdust\ -- which is a completely independent quantity. However, we acknowledge the current limited sample size and the paucity of line variability measurements. More complete sampling of AGN in both X-rays and in the infrared are clearly important for drawing robust conclusions, and we discuss future possibilities in the next section.

\subsection{Future perspectives}

This field is expected to leap forward with the imminent launch of the \astroh\ mission \citep{astroh14}. With a spectral resolution of $\approx$\,4--7\,eV, the Soft X-ray Spectrometer (SXS) employing calorimetric photon detection will provide an improvement in spectral resolution of $\approx$\,6--10 as compared to present grating instruments. Since \rfe\ has an inverse quadratic scaling with line velocity, SXS will extend the radius out to which the K$\alpha$ emission can be localized by a factor of $\gtsim$\,40 in any individual object. This assumes a single Gaussian feature. Individual ionization stages will not all be necessarily separable by the SXS, though some deconvolution will be possible. In terms of the overall sensitivity to line parameter measurement, this can be quantified in terms of a \lq figure of merit\rq, which combines the effective area and the spectral resolution. Relative to HEG, the overall SXS sensitivity to line detection is expected to be improved by a factor of $\approx$\,7, extending our reach to correspondingly faint (and more distant) systems. However, the improvement in the corresponding figure of merit for detection of line {\em broadening} (which is key to our work) is about two orders of magnitude relative to the HEG.\footnote{{\tt http://astro-h.isas.jaxa.jp/ahqr.pdf}}

In the next decade, the \athena\ mission \citep{athena} will have a much larger collecting area than \astroh, pushing such studies out to high redshift. Reverberation of the narrow \fek\ core for more objects would also be an independent constraint on \rfe. Variable narrow lines in some cases have already been mentioned. In addition, \citet{liu10} found Fe line reverberation on a timescale of $\sim$\,20--40 day in NGC\,5548. This would push \rfe\ a factor of $\sim$\,1.5--3 lower than our estimate in Table\,1, but still consistent within the uncertainty on \rfe. 

And in a similar vein, enlarged NIR coverage will greatly help to better fill the \rfe\ vs. \rdust\ parameter space. There are currently a total of 31 AGN with measurements of \rdust. This includes sources from the references cited in Table\,1 as well as a few other published and unpublished sources (e.g. GQ\,Com; \citealt{sitko93}). We have included 13 of these (42\,\%) in our X-ray cross-matched sample. Although Keck Interferometer is no longer available, more size measurements will be available soon (Kishimoto et al. 2015 submitted) making use of the three-telescope beam combiner AMBER at the VLTI. This will boost the present sample and address projection effects by invoking closure phase data. And although the MAGNUM telescope is now decommissioned, there are many other world-wide efforts ongoing to obtain AGN NIR time lags \citep[e.g. ][]{pozonunez15}. LSST could also serve as a dust reverberation machine providing thousands of time lag measurements \citep{hoenig14_lsst}. Finally, in the distant future, we can expect to directly resolve the torus in many more sources with larger interferometers such as the Planet Formation Imager \citep{pfi}.

\vspace*{1cm}
\acknowledgements
\noindent
We acknowledge funding from STFC (ST/J003697/1) for P.G., a Marie Curie International Incoming Fellowship within the 7$^{\rm th}$ EC Framework (PIIF-GA-2013-623804) for S.F.H. and JSPS (26887044) for M.K. We thank C.\,Done, Y.\,Ueda, R.F.\,Mushotzky, D.R.\,Ballantyne, J.H. Krolik and R. Antonucci for comments on an initial draft which greatly improved the discussion of the robustness of our results. We also acknowledge useful comments from the anonymous referee on the first submission and the revised draft.

\clearpage

\begin{sidewaystable}
\begin{center}
{\scriptsize
\caption{Sample properties\label{tab:sample}}
\begin{tabular}{lcccccccccr}
\hline
\hline
Source & $M_{\rm BH}$                   & Ref.($M_{\rm BH}$) & $\tau_{\rm dust}$   & $R_{\rm dust,rev}$     & $R_{\rm dust,intf}$ & Ref.($R_{\rm dust}$) & $R_{\rm BLR}$      & $L_{\rm Bol}$ & $v_{\rm FWHM}^{\rm Fe}$ & $R_{\rm Fe}$            \\
       &   $\times$\,10$^6$\,M$_\odot$  &                  & days              & $\times$\,0.1\,pc  & $\times$\,0.1\,pc &                   & $\times$0.01\,pc & 10$^{45}$\,erg\,s$^{-1}$ & km\,s$^{-1}$         & $\times$\,0.1\,pc     \\
\hline
Fairall9 & 255.0$\pm$56.0 & i & 400.0\,$\pm$\,100.0 & 3.21$\pm$\,0.80 & -- &        1 & 1.40$_{-0.35}^{+0.26}$ & 1.90 & 18100$^{+76840}_{-12390}$ & 0.045$^{+0.135}_{-0.045}$ \\
Mrk590 & 15.8$\pm$3.4 & ii & 32.2\,$\pm$\,4.0 & 0.26$\pm$\,0.03 & -- &        2 & 2.09$_{-0.19}^{0.16}$  & 0.17 & 4350$^{+6060}_{-2030}$ & 0.048$^{+0.077}_{-0.045}$ \\
NGC3516 & 42.7$\pm$14.6 & i & 71.5\,$\pm$\,5.8 & 0.59$\pm$\,0.05 & -- &        2 & 0.56$_{-0.32}^{0.57}$  & 0.025 & 3180$^{+880}_{-670}$ & 0.242$^{+0.177}_{-0.113}$ \\
NGC3783 & 29.8$\pm$5.4 & i & 76.3\,$^{+10.9}_{-17.2}$ & 0.64$^{+0.09}_{-0.14}$ & 1.60\,$\pm$\,0.50 &        3;        4 & 0.84$_{-0.19}^{+0.27}$ & 0.17 & 1750$^{+360}_{-360}$ & 0.558$^{+0.314}_{-0.203}$ \\
NGC4051 & 1.9$\pm$0.8 & i & 15.8\,$\pm$\,0.5 & 0.13$\pm$\,0.004 & 0.32\,$\pm$\,0.05 &        2;        5 & 0.49$_{-0.15}^{+0.22}$ & 0.0059 & $<$\,1860 & $>$\,0.032 \\
NGC4151 & 48.5$\pm$20.0 & iii & 36.0\,$^{+9.0}_{-7.0}$ & 0.30$^{+0.08}_{-0.06}$ & 0.41\,$\pm$\,0.04 &        6;        5 & 0.55$_{-0.07}^{+0.09}$ & 0.050 & 2250$^{+400}_{-360}$ & 0.549$^{+0.377}_{-0.228}$ \\
3C273 & 6590.0$\pm$1300.0 & iv & -- & -- & 8.80\,$\pm$\,4.70 &        7 & 22.23$_{-6.59}^{+4.96}$ & 58.80 & 5900$^{+8640}_{-5830}$ & 10.857$^{+67.554}_{-10.240}$ \\
NGC4593 & 5.4$\pm$9.4 & i & 42.1\,$\pm$\,0.9 & 0.35$\pm$\,0.01 & -- &        2 & 0.31$_{-0.07}^{+0.07}$ & 0.042 & 2230$^{+8180}_{-1100}$ & 0.062$^{+0.346}_{-0.062}$ \\
IRAS13349+2438 & 1000.0$\pm$1000.0 & v & -- & -- & 8.20\,$\pm$\,3.40 &        8 & -- & 11.16 & 5150$^{+60200}_{-2810}$ & 2.162$^{+7.217}_{-2.162}$ \\
IC4329A & 9.9$\pm$17.9 & i & -- & -- & 2.90\,$\pm$\,0.40 &        8 & 0.12$_{-0.15}^{+0.22}$ & 0.79 & $<$\,1860 & $>$\,0.164 \\
NGC5548 & 67.1$\pm$2.6 & i & 61.3\,$\pm$\,0.3 & 0.51$\pm$\,0.002 & -- &        2 & 1.49$_{-0.05}^{+0.05}$ & 0.07 & 2540$^{+1140}_{-820}$ & 0.596$^{+0.545}_{-0.358}$ \\
Mrk509 & 143.0$\pm$12.0 & i & 120.3\,$\pm$\,1.1 & 0.98$\pm$\,0.01 & 3.00\,$\pm$\,0.50 &        2;        9 & 6.46$_{-0.44}^{+0.50}$ & 1.04 & 2910$^{+2590}_{-1250}$ & 0.968$^{+1.337}_{-0.801}$ \\
NGC7469 & 12.2$\pm$1.4 & i & 78.1\,$\pm$\,0.1 & 0.65$\pm$\,0.0004 & $<$\,1.90 &        2;        8 & 0.37$_{-0.07}^{+0.06}$ & 0.19 & 4890$^{+2770}_{-1700}$ & 0.029$^{+0.029}_{-0.020}$ \\
\hline
\end{tabular}
}
\end{center}

~\par
Notes: Uncertainties are quoted for 68\,\%\ confidence. $M_{\rm BH}$ ref.: i: \citet{peterson04}; ii: \citet{kaspi00}; iii: \citet{hoenig14_4151}; iv: \citet{paltani05}; v: \citet{lee13}. $R_{\rm dust}$ ref.: 1: \citet{clavel89}; 2: \citet{koshida14}; 3: \citet{lira11}; 4: \citet{weigelt12}; 5: \citet{kishimoto09}; 6: \citet{hoenig14_4151}; 7: \citet{kishimoto11}; 8: Kishimoto et al. (2015, submitted); 9: \citet{kishimoto13}.\\
\end{sidewaystable}

\appendix

\section{Black hole masses independent of reverberation mapping}

Our adopted values of \mbh\ in Table\,1 are mostly virial estimates based upon reverberation mapping of the BLR, which implicitly assume a certain correction for geometric projection effects. The size comparisons in Fig.\,1 effectively presuppose that the same projection effects apply for the \fek\ emission regions and the BLR, which may, or may not, be the case. Although the correction factors are expected to be small (see Section\,\ref{sec:xraysample}), it is worth investigating how strongly the above assumption affects our results. 
In order to assess this, we compile black hole masses which are independent of reverberation mapping. Any distinct projection effect for the \fek\ and \hb\ emission regions will then be retained in the emitter size comparisons, and the resultant change in the \rfe\ distribution (relative to Fig.\,1) will give an estimate of the maximal impact that the unknown projection effects can have. 

Table\,\ref{tab:msigma} presents our compilation of alternate \mbh\ values.\footnote{Although our primary adopted value of \mbh\,(NGC\,4151) from \citet{hoenig14_4151} is not based upon reverberation mapping, we use an alternate \mbh\ value based upon the \mbh--$\sigma_*$ method here as a cross-check.} These are mostly based upon host galaxy stellar velocity dispersion ($\sigma_*$) measurements and the relation between \mbh\ and $\sigma_*$ from \citet{gultekin09}. Where this was not possible, we used scaling relations between X-ray variability amplitude (\citealt{zhou10, mchardy13}) and \mbh, or the empirical relation between optical continuum luminosity and the BLR radius \citep{vestergaard02} combined with the FWHM of the \hb\ emission line. There is broad agreement between the alternate masses tabulated here and those listed in Table\,1, with the alternate masses (mostly based upon \mbh--$\sigma_*$) being larger than the virial estimates by a median factor of 1.8 but with a significant spread of 0.7 dex (standard deviation) between the two sets of masses. 

\begin{figure*}
  \begin{center}
  \includegraphics[angle=90,width=13cm]{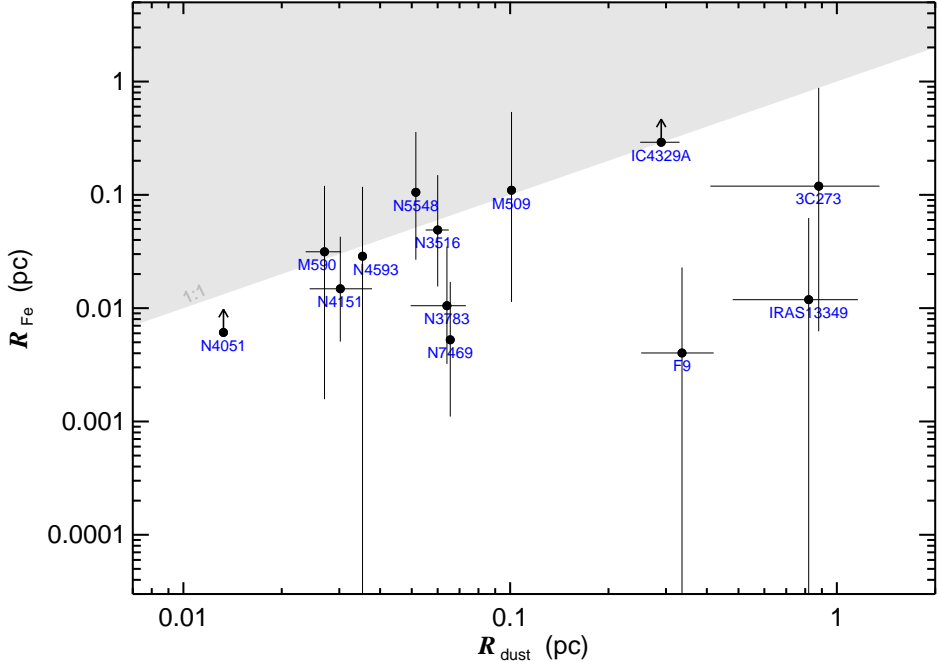}
  \caption{The comparison of \rfe\ with \rreverb\ as in Fig.\,1, but this time with the computation of \rfe\ based upon black hole masses obtained from reverberation-independent methods.
    \label{fig:compare_sizes_msigma}}
  \end{center}
\end{figure*}

The resultant values of \rfe\ are also listed in the table, and are plotted in Fig.\,\ref{fig:compare_sizes_msigma} which shows that although the location of the \fek\ emission radii relative to \rdust\ (and thus also relative to \rblr) do change in some individual objects as compared to Fig.\,1 (e.g. \rfe(NGC\,3783) is now smaller than \rdust, and \rfe(Mrk\,590) is now consistent with \rdust), our main inference of \rdust\ serving as an approximate envelope to \rfe\ still holds. IC\,4329A is the only source in which the limit to \rfe\ apparently sits very close to (but still consistent with) \rdust. As we discussed in Section\,\ref{sec:discussion1}, \fek\ emission will not cease exactly at \rdust, and \rfe\ values of up to a few times \rdust\ are allowed by Fig.\,1 and also by typical compact torus models. \astroh\ observations of this source will help to pinpoint its location on the \rfe--\rdust\ plane.

\begin{table}
\begin{center}
{\scriptsize
\caption{Alternate black hole mass estimates\label{tab:msigma}}
\begin{tabular}{lcccccr}
\hline
\hline
Source          & Method(\mbh)      & Measurement    & Ref.           & log\,$M_{\rm BH}$  &  $\Delta$log\,$M_{\rm BH}$  & $R_{\rm Fe}$            \\
                 &                  &  &                  & \Msun            & \Msun           & $\times$\,0.1\,pc     \\
(1)              &   (2)            &  (3)&  (4)          & (5)              & (6)             & (7)                   \\
\hline                                                              
Fairall\,9      & \mbh--$\sigma_*$  & $\sigma_*$=228\p18\,km\,s$^{-1}$     & \citet{oliva99}  & 8.36\p0.46 & +0.05\p0.47& 0.040$^{+0.186}_{-0.040}$ \\
Mrk\,590        & \mbh--$\sigma_*$  & $\sigma_*$=189\p6\,km\,s$^{-1}$      & \citet{nelson04} & 8.02\p0.44 &--0.82\p0.45& 0.314$^{+0.890}_{-0.298}$ \\
NGC\,3516       & \mbh--$\sigma_*$  & $\sigma_*$=181\p5\,km\,s$^{-1}$      & \citet{nelson04} & 7.94\p0.44 &--0.31\p0.47& 0.490$^{+1.016}_{-0.335}$ \\
NGC\,3783       & \mbh--$\sigma_*$  & $\sigma_*$=95\p10\,km\,s$^{-1}$      & \citet{onken04}  & 6.75\p0.48 & +0.73\p0.49& 0.105$^{+0.241}_{-0.073}$ \\
NGC\,4051       & \mbh--$\sigma_*$  & $\sigma_*$=86\p3\,km\,s$^{-1}$       & \citet{nelson04} & 6.57\p0.44 &--0.28\p0.47& $>$\,0.061 \\
NGC\,4151       & \mbh--$\sigma_*$  & $\sigma_*$=116\p3\,km\,s$^{-1}$      & \citet{onken14}  & 7.12\p0.44 & +0.57\p0.48& 0.148$^{+0.288}_{-0.098}$ \\
3C\,273         & X-ray variance    & log(NVA)\,=\,0.20\p0.08             & \citet{mchardy13}& 8.86\p0.08 & +0.96\p0.12&  1.193$^{+7.775}_{-1.129}$ \\
NGC\,4593       & \mbh--$\sigma_*$  & $\sigma_*$=135\p6\,km\,s$^{-1}$      & \citet{nelson04} & 7.40\p0.45 &--0.67\p0.88& 0.287$^{+0.903}_{-0.287}$ \\
IRAS\,13349+2438& $L_{5100}$--\rblr & $L_{5100}$=10$^{44.64}$\,erg\,s$^{-1}$ & \citet{wang04, grupe04} & 7.74\p0.50 &+1.26\p0.66& 0.119$^{+0.474}_{-0.119}$ \\
IC\,4329A       & \mbh--$\sigma_*$  & $\sigma_*$=225\p9\,km\,s$^{-1}$      & \citet{oliva99}  & 8.24\p0.44 &--1.25\p0.90& $>$\,2.911 \\
NGC\,5548       & \mbh--$\sigma_*$  & $\sigma_*$=195\p13\,km\,s$^{-1}$     & \citet{woo10}    & 8.07\p0.46 &--0.25\p0.46& 1.053$^{+2.659}_{-0.788}$ \\
Mrk\,509        & X-ray variance    & log($\sigma_{\rm rms}^2$)=--3.24\p0.59& \citet{zhou10}   & 8.21\p0.62 &--0.05\p0.62& 1.098$^{+4.413}_{-0.993}$ \\
NGC\,7469       & \mbh--$\sigma_*$  & $\sigma_*$=131\p5\,km\,s$^{-1}$      & \citet{nelson04} & 7.34\p0.45 &--0.25\p0.45& 0.053$^{+0.122}_{-0.042}$ \\
\hline
\end{tabular}
~\par
Column (2) denotes the method used to determine \mbh. Three methods are used: the \mbh--$\sigma_*$\ relation from \citet{gultekin09}; the X-ray variance scaling with \mbh\ either from \citet[][who characterizes this in terms of the Normalized Variability Amplitude or NVA]{mchardy13} or from \citet[][who provide measurements of the rms variability $\sigma_{\rm rms}^2$]{zhou10}; the empirical relation of $L_{5100}$ with \rblr\ from \citet{vestergaard02}. Column (3) lists the relevant observable used in the method. Column (4) gives the reference for the measurement of the observable. Column (5) lists the resultant \mbh\ value, and Column (6) the difference in log\,\mbh\ with respect to the value quoted and used in the main paper (Table\,1). Finally, Column (7) lists the corresponding value of \rfe\ using the alternate \mbh\ estimate.  For IRAS\,13349+2438, we assume a large error of 0.5\,dex in \mbh\ as in \citet{demarco13}. Note that for this source $v_{\rm FWHM}^{\rm H\beta}$=2800\,\p\,180\,km\,s$^{-1}$ is reported in \citet{grupe04}.
}
\end{center}
\end{table}

\bibliographystyle{apj.bst}

\begin{thebibliography}{78}
\expandafter\ifx\csname natexlab\endcsname\relax\def\natexlab#1{#1}\fi

\bibitem[{{Ar{\'e}valo} {et~al.}(2014){Ar{\'e}valo}, {Bauer}, {Puccetti},
  {Walton}, {Koss}, {Boggs}, {Brandt}, {Brightman}, {Christensen}, {Comastri},
  {Craig}, {Fuerst}, {Gandhi}, {Grefenstette}, {Hailey}, {Harrison}, {Luo},
  {Madejski}, {Madsen}, {Marinucci}, {Matt}, {Saez}, {Stern}, {Stuhlinger},
  {Treister}, {Urry}, \& {Zhang}}]{arevalo14}
{Ar{\'e}valo}, P., {Bauer}, F.~E., {Puccetti}, S., {Walton}, D.~J., {Koss}, M.,
  {Boggs}, S.~E., {Brandt}, W.~N., {Brightman}, M., {Christensen}, F.~E.,
  {Comastri}, A., {Craig}, W.~W., {Fuerst}, F., {Gandhi}, P., {Grefenstette},
  B.~W., {Hailey}, C.~J., {Harrison}, F.~A., {Luo}, B., {Madejski}, G.,
  {Madsen}, K.~K., {Marinucci}, A., {Matt}, G., {Saez}, C., {Stern}, D.,
  {Stuhlinger}, M., {Treister}, E., {Urry}, C.~M., \& {Zhang}, W.~W. 2014,
  \apj, 791, 81

\bibitem[{{Barvainis}(1987)}]{barvainis87}
{Barvainis}, R. 1987, \apj, 320, 537

\bibitem[{{Bauer} {et~al.}(2014){Bauer}, {Arevalo}, {Walton}, {Koss},
  {Puccetti}, {Gandhi}, {Stern}, {Alexander}, {Balokovic}, {Boggs}, {Brandt},
  {Brightman}, {Christensen}, {Comastri}, {Craig}, {Del Moro}, {Hailey},
  {Harrison}, {Hickox}, {Luo}, {Markwardt}, {Marinucci}, {Matt}, {Rigby},
  {Rivers}, {Saez}, {Treister}, {Urry}, \& {Zhang}}]{bauer15}
{Bauer}, F.~E., {Arevalo}, P., {Walton}, D.~J., {Koss}, M.~J., {Puccetti}, S.,
  {Gandhi}, P., {Stern}, D., {Alexander}, D.~M., {Balokovic}, M., {Boggs},
  S.~E., {Brandt}, W.~N., {Brightman}, M., {Christensen}, F.~E., {Comastri},
  A., {Craig}, W.~W., {Del Moro}, A., {Hailey}, C.~J., {Harrison}, F.~A.,
  {Hickox}, R., {Luo}, B., {Markwardt}, C.~B., {Marinucci}, A., {Matt}, G.,
  {Rigby}, J.~R., {Rivers}, E., {Saez}, C., {Treister}, E., {Urry}, C.~M., \&
  {Zhang}, W.~W. 2014, ApJ submitted, arXiv:1411.0670

\bibitem[{{Bentz} {et~al.}(2009){Bentz}, {Peterson}, {Netzer}, {Pogge}, \&
  {Vestergaard}}]{bentz09}
{Bentz}, M.~C., {Peterson}, B.~M., {Netzer}, H., {Pogge}, R.~W., \&
  {Vestergaard}, M. 2009, \apj, 697, 160

\bibitem[{{Bianchi} {et~al.}(2008){Bianchi}, {La Franca}, {Matt}, {Guainazzi},
  {Jimenez Bail{\'o}n}, {Longinotti}, {Nicastro}, \&
  {Pentericci}}]{bianchi08_n7213}
{Bianchi}, S., {La Franca}, F., {Matt}, G., {Guainazzi}, M., {Jimenez
  Bail{\'o}n}, E., {Longinotti}, A.~L., {Nicastro}, F., \& {Pentericci}, L.
  2008, \mnras, 389, L52

\bibitem[{{Brenneman} {et~al.}(2014){Brenneman}, {Madejski}, {Fuerst}, {Matt},
  {Elvis}, {Harrison}, {Ballantyne}, {Boggs}, {Christensen}, {Craig}, {Fabian},
  {Grefenstette}, {Hailey}, {Madsen}, {Marinucci}, {Rivers}, {Stern}, {Walton},
  \& {Zhang}}]{brenneman14}
{Brenneman}, L.~W., {Madejski}, G., {Fuerst}, F., {Matt}, G., {Elvis}, M.,
  {Harrison}, F.~A., {Ballantyne}, D.~R., {Boggs}, S.~E., {Christensen}, F.~E.,
  {Craig}, W.~W., {Fabian}, A.~C., {Grefenstette}, B.~W., {Hailey}, C.~J.,
  {Madsen}, K.~K., {Marinucci}, A., {Rivers}, E., {Stern}, D., {Walton}, D.~J.,
  \& {Zhang}, W.~W. 2014, \apj, 788, 61

\bibitem[{{Brightman} \& {Nandra}(2011)}]{brightmannandra11}
{Brightman}, M., \& {Nandra}, K. 2011, \mnras, 413, 1206

\bibitem[{{Burtscher} {et~al.}(2013){Burtscher}, {Meisenheimer}, {Tristram},
  {Jaffe}, {H{\"o}nig}, {Davies}, {Kishimoto}, {Pott}, {R{\"o}ttgering},
  {Schartmann}, {Weigelt}, \& {Wolf}}]{burtscher13}
{Burtscher}, L., {Meisenheimer}, K., {Tristram}, K.~R.~W., {Jaffe}, W.,
  {H{\"o}nig}, S.~F., {Davies}, R.~I., {Kishimoto}, M., {Pott}, J.-U.,
  {R{\"o}ttgering}, H., {Schartmann}, M., {Weigelt}, G., \& {Wolf}, S. 2013,
  \aap, 558, A149

\bibitem[{{Clavel} {et~al.}(1989){Clavel}, {Wamsteker}, \& {Glass}}]{clavel89}
{Clavel}, J., {Wamsteker}, W., \& {Glass}, I.~S. 1989, \apj, 337, 236

\bibitem[{{De Marco} {et~al.}(2013){De Marco}, {Ponti}, {Cappi}, {Dadina},
  {Uttley}, {Cackett}, {Fabian}, \& {Miniutti}}]{demarco13}
{De Marco}, B., {Ponti}, G., {Cappi}, M., {Dadina}, M., {Uttley}, P.,
  {Cackett}, E.~M., {Fabian}, A.~C., \& {Miniutti}, G. 2013, \mnras, 431, 2441

\bibitem[{{Fabian} {et~al.}(1989){Fabian}, {Rees}, {Stella}, \&
  {White}}]{fabian89}
{Fabian}, A.~C., {Rees}, M.~J., {Stella}, L., \& {White}, N.~E. 1989, \mnras,
  238, 729

\bibitem[{{Gandhi} {et~al.}(2009){Gandhi}, {Horst}, {Smette}, {H{\"o}nig},
  {Comastri}, {Gilli}, {Vignali}, \& {Duschl}}]{g09_mirxray}
{Gandhi}, P., {Horst}, H., {Smette}, A., {H{\"o}nig}, S., {Comastri}, A.,
  {Gilli}, R., {Vignali}, C., \& {Duschl}, W. 2009, \aap, 502, 457

\bibitem[{{Gandhi} {et~al.}(2014){Gandhi}, {Lansbury}, {Alexander}, {Stern},
  {Ar{\'e}valo}, {Ballantyne}, {Balokovi{\'c}}, {Bauer}, {Boggs}, {Brandt},
  {Brightman}, {Christensen}, {Comastri}, {Craig}, {Del Moro}, {Elvis},
  {Fabian}, {Hailey}, {Harrison}, {Hickox}, {Koss}, {LaMassa}, {Luo},
  {Madejski}, {Ptak}, {Puccetti}, {Teng}, {Urry}, {Walton}, \& {Zhang}}]{g14}
{Gandhi}, P., {Lansbury}, G.~B., {Alexander}, D.~M., {Stern}, D.,
  {Ar{\'e}valo}, P., {Ballantyne}, D.~R., {Balokovi{\'c}}, M., {Bauer}, F.~E.,
  {Boggs}, S.~E., {Brandt}, W.~N., {Brightman}, M., {Christensen}, F.~E.,
  {Comastri}, A., {Craig}, W.~W., {Del Moro}, A., {Elvis}, M., {Fabian}, A.~C.,
  {Hailey}, C.~J., {Harrison}, F.~A., {Hickox}, R.~C., {Koss}, M., {LaMassa},
  S.~M., {Luo}, B., {Madejski}, G.~M., {Ptak}, A.~F., {Puccetti}, S., {Teng},
  S.~H., {Urry}, C.~M., {Walton}, D.~J., \& {Zhang}, W.~W. 2014, \apj, 792, 117

\bibitem[{{Gandhi} {et~al.}(2015){Gandhi}, {Yamada}, {Ricci}, {Asmus},
  {Mushotzky}, {Ueda}, {Terashima}, \& {La Parola}}]{g15_n4785}
{Gandhi}, P., {Yamada}, S., {Ricci}, C., {Asmus}, D., {Mushotzky}, R.~F.,
  {Ueda}, Y., {Terashima}, Y., \& {La Parola}, V. 2015, \mnras, 449, 1845

\bibitem[{{George} \& {Fabian}(1991)}]{georgefabian91}
{George}, I.~M., \& {Fabian}, A.~C. 1991, \mnras, 249, 352

\bibitem[{{Gibson} {et~al.}(2005){Gibson}, {Marshall}, {Canizares}, \&
  {Lee}}]{gibson05}
{Gibson}, R.~R., {Marshall}, H.~L., {Canizares}, C.~R., \& {Lee}, J.~C. 2005,
  \apj, 627, 83

\bibitem[{{Grupe} {et~al.}(2004){Grupe}, {Wills}, {Leighly}, \&
  {Meusinger}}]{grupe04}
{Grupe}, D., {Wills}, B.~J., {Leighly}, K.~M., \& {Meusinger}, H. 2004, \aj,
  127, 156

\bibitem[{{G{\"u}ltekin} {et~al.}(2009){G{\"u}ltekin}, {Richstone}, {Gebhardt},
  {Lauer}, {Tremaine}, {Aller}, {Bender}, {Dressler}, {Faber}, {Filippenko},
  {Green}, {Ho}, {Kormendy}, {Magorrian}, {Pinkney}, \& {Siopis}}]{gultekin09}
{G{\"u}ltekin}, K., {Richstone}, D.~O., {Gebhardt}, K., {Lauer}, T.~R.,
  {Tremaine}, S., {Aller}, M.~C., {Bender}, R., {Dressler}, A., {Faber}, S.~M.,
  {Filippenko}, A.~V., {Green}, R., {Ho}, L.~C., {Kormendy}, J., {Magorrian},
  J., {Pinkney}, J., \& {Siopis}, C. 2009, \apj, 698, 198

\bibitem[{{Hagino} {et~al.}(2015){Hagino}, {Odaka}, {Done}, {Gandhi},
  {Watanabe}, {Sako}, \& {Takahashi}}]{hagino15}
{Hagino}, K., {Odaka}, H., {Done}, C., {Gandhi}, P., {Watanabe}, S., {Sako},
  M., \& {Takahashi}, T. 2015, \mnras, 446, 663

\bibitem[{{H{\"o}nig}(2014)}]{hoenig14_lsst}
{H{\"o}nig}, S.~F. 2014, \apjl, 784, L4

\bibitem[{{H{\"o}nig} \& {Kishimoto}(2011)}]{hoenig11_variability}
{H{\"o}nig}, S.~F., \& {Kishimoto}, M. 2011, \aap, 534, A121

\bibitem[{{H{\"o}nig} {et~al.}(2014){H{\"o}nig}, {Watson}, {Kishimoto}, \&
  {Hjorth}}]{hoenig14_4151}
{H{\"o}nig}, S.~F., {Watson}, D., {Kishimoto}, M., \& {Hjorth}, J. 2014, \nat,
  515, 528

\bibitem[{{Jiang} {et~al.}(2011){Jiang}, {Wang}, \& {Shu}}]{jiang11_fek}
{Jiang}, P., {Wang}, J., \& {Shu}, X. 2011, Science China Physics, Mechanics,
  and Astronomy, 54, 1354

\bibitem[{{Kaspi} {et~al.}(2000){Kaspi}, {Smith}, {Netzer}, {Maoz}, {Jannuzi},
  \& {Giveon}}]{kaspi00}
{Kaspi}, S., {Smith}, P.~S., {Netzer}, H., {Maoz}, D., {Jannuzi}, B.~T., \&
  {Giveon}, U. 2000, \apj, 533, 631

\bibitem[{{Kishimoto} {et~al.}(2011{\natexlab{a}}){Kishimoto}, {H{\"o}nig},
  {Antonucci}, {Barvainis}, {Kotani}, {Tristram}, {Weigelt}, \&
  {Levin}}]{kishimoto11}
{Kishimoto}, M., {H{\"o}nig}, S.~F., {Antonucci}, R., {Barvainis}, R.,
  {Kotani}, T., {Tristram}, K.~R.~W., {Weigelt}, G., \& {Levin}, K.
  2011{\natexlab{a}}, \aap, 527, A121

\bibitem[{{Kishimoto} {et~al.}(2009){Kishimoto}, {H{\"o}nig}, {Antonucci},
  {Kotani}, {Barvainis}, {Tristram}, \& {Weigelt}}]{kishimoto09}
{Kishimoto}, M., {H{\"o}nig}, S.~F., {Antonucci}, R., {Kotani}, T.,
  {Barvainis}, R., {Tristram}, K.~R.~W., \& {Weigelt}, G. 2009, \aap, 507, L57

\bibitem[{{Kishimoto} {et~al.}(2013){Kishimoto}, {H{\"o}nig}, {Antonucci},
  {Millan-Gabet}, {Barvainis}, {Millour}, {Kotani}, {Tristram}, \&
  {Weigelt}}]{kishimoto13}
{Kishimoto}, M., {H{\"o}nig}, S.~F., {Antonucci}, R., {Millan-Gabet}, R.,
  {Barvainis}, R., {Millour}, F., {Kotani}, T., {Tristram}, K.~R.~W., \&
  {Weigelt}, G. 2013, \apjl, 775, L36

\bibitem[{{Kishimoto} {et~al.}(2011{\natexlab{b}}){Kishimoto}, {H{\"o}nig},
  {Antonucci}, {Millour}, {Tristram}, \& {Weigelt}}]{kishimoto11b}
{Kishimoto}, M., {H{\"o}nig}, S.~F., {Antonucci}, R., {Millour}, F.,
  {Tristram}, K.~R.~W., \& {Weigelt}, G. 2011{\natexlab{b}}, \aap, 536, A78

\bibitem[{{Koshida} {et~al.}(2014){Koshida}, {Minezaki}, {Yoshii}, {Kobayashi},
  {Sakata}, {Sugawara}, {Enya}, {Suganuma}, {Tomita}, {Aoki}, \&
  {Peterson}}]{koshida14}
{Koshida}, S., {Minezaki}, T., {Yoshii}, Y., {Kobayashi}, Y., {Sakata}, Y.,
  {Sugawara}, S., {Enya}, K., {Suganuma}, M., {Tomita}, H., {Aoki}, T., \&
  {Peterson}, B.~A. 2014, \apj, 788, 159

\bibitem[{{Krolik} \& {Kallman}(1987)}]{krolik87}
{Krolik}, J.~H., \& {Kallman}, T.~R. 1987, \apjl, 320, L5

\bibitem[{{Lee} {et~al.}(2013){Lee}, {Kriss}, {Chakravorty}, {Rahoui}, {Young},
  {Brandt}, {Hines}, {Ogle}, \& {Reynolds}}]{lee13}
{Lee}, J.~C., {Kriss}, G.~A., {Chakravorty}, S., {Rahoui}, F., {Young}, A.~J.,
  {Brandt}, W.~N., {Hines}, D.~C., {Ogle}, P.~M., \& {Reynolds}, C.~S. 2013,
  \mnras, 430, 2650

\bibitem[{{Lira} {et~al.}(2011){Lira}, {Ar{\'e}valo}, {Uttley}, {McHardy}, \&
  {Breedt}}]{lira11}
{Lira}, P., {Ar{\'e}valo}, P., {Uttley}, P., {McHardy}, I., \& {Breedt}, E.
  2011, \mnras, 415, 1290

\bibitem[{{Liu} {et~al.}(2010){Liu}, {Elvis}, {McHardy}, {Grupe}, {Wilkes},
  {Reeves}, {Brickhouse}, {Krongold}, {Mathur}, {Minezaki}, {Nicastro},
  {Yoshii}, \& {Zhang}}]{liu10}
{Liu}, Y., {Elvis}, M., {McHardy}, I.~M., {Grupe}, D., {Wilkes}, B.~J.,
  {Reeves}, J., {Brickhouse}, N., {Krongold}, Y., {Mathur}, S., {Minezaki}, T.,
  {Nicastro}, F., {Yoshii}, Y., \& {Zhang}, S.~N. 2010, \apj, 710, 1228

\bibitem[{{Lobban} {et~al.}(2011){Lobban}, {Reeves}, {Miller}, {Turner},
  {Braito}, {Kraemer}, \& {Crenshaw}}]{lobban11}
{Lobban}, A.~P., {Reeves}, J.~N., {Miller}, L., {Turner}, T.~J., {Braito}, V.,
  {Kraemer}, S.~B., \& {Crenshaw}, D.~M. 2011, \mnras, 414, 1965

\bibitem[{{Marinucci} {et~al.}(2015){Marinucci}, {Matt}, {Bianchi}, {Lu},
  {Arevalo}, {Balokovi{\'c}}, {Ballantyne}, {Bauer}, {Boggs}, {Christensen},
  {Craig}, {Gandhi}, {Hailey}, {Harrison}, {Puccetti}, {Rivers}, {Walton},
  {Stern}, \& {Zhang}}]{marinucci15}
{Marinucci}, A., {Matt}, G., {Bianchi}, S., {Lu}, T.~N., {Arevalo}, P.,
  {Balokovi{\'c}}, M., {Ballantyne}, D., {Bauer}, F.~E., {Boggs}, S.~E.,
  {Christensen}, F.~E., {Craig}, W.~W., {Gandhi}, P., {Hailey}, C.~J.,
  {Harrison}, F., {Puccetti}, S., {Rivers}, E., {Walton}, D.~J., {Stern}, D.,
  \& {Zhang}, W. 2015, \mnras, 447, 160

\bibitem[{{Marinucci} {et~al.}(2012){Marinucci}, {Risaliti}, {Wang}, {Nardini},
  {Elvis}, {Fabbiano}, {Bianchi}, \& {Matt}}]{marinucci12}
{Marinucci}, A., {Risaliti}, G., {Wang}, J., {Nardini}, E., {Elvis}, M.,
  {Fabbiano}, G., {Bianchi}, S., \& {Matt}, G. 2012, \mnras, 423, L6

\bibitem[{{Matt}(2002)}]{matt02}
{Matt}, G. 2002, \mnras, 337, 147

\bibitem[{{McHardy}(2013)}]{mchardy13}
{McHardy}, I.~M. 2013, \mnras, 430, L49

\bibitem[{{McKernan} \& {Yaqoob}(2004)}]{mckernan04}
{McKernan}, B., \& {Yaqoob}, T. 2004, \apj, 608, 157

\bibitem[{{Mendoza} {et~al.}(2004){Mendoza}, {Kallman}, {Bautista}, \&
  {Palmeri}}]{mendoza04}
{Mendoza}, C., {Kallman}, T.~R., {Bautista}, M.~A., \& {Palmeri}, P. 2004,
  \aap, 414, 377

\bibitem[{{Minezaki} \& {Matsushita}(2015)}]{minezaki15}
{Minezaki}, T., \& {Matsushita}, K. 2015, \apj, 802, 98

\bibitem[{{Monnier} {et~al.}(2014){Monnier}, {Kraus}, {Buscher}, {Berger},
  {Haniff}, {Ireland}, {Labadie}, {Lacour}, {Le Coroller}, {Petrov}, {Pott},
  {Ridgway}, {Surdej}, {ten Brummelaar}, {Tuthill}, \& {van Belle}}]{pfi}
{Monnier}, J.~D., {Kraus}, S., {Buscher}, D., {Berger}, J.-P., {Haniff}, C.,
  {Ireland}, M., {Labadie}, L., {Lacour}, S., {Le Coroller}, H., {Petrov},
  R.~G., {Pott}, J.-U., {Ridgway}, S., {Surdej}, J., {ten Brummelaar}, T.,
  {Tuthill}, P., \& {van Belle}, G. 2014, in Society of Photo-Optical
  Instrumentation Engineers (SPIE) Conference Series, Vol. 9146, Society of
  Photo-Optical Instrumentation Engineers (SPIE) Conference Series, 10

\bibitem[{{Murphy} \& {Yaqoob}(2009)}]{mytorus}
{Murphy}, K.~D., \& {Yaqoob}, T. 2009, \mnras, 397, 1549

\bibitem[{{Nandra}(2006)}]{nandra06}
{Nandra}, K. 2006, \mnras, 368, L62

\bibitem[{{Nandra} {et~al.}(2013){Nandra}, {Barret}, {Barcons}, {Fabian}, {den
  Herder}, {Piro}, {Watson}, {Adami}, {Aird}, {Afonso}, \& et~al.}]{athena}
{Nandra}, K., {Barret}, D., {Barcons}, X., {Fabian}, A., {den Herder}, J.-W.,
  {Piro}, L., {Watson}, M., {Adami}, C., {Aird}, J., {Afonso}, J.~M., \& et~al.
  2013, The Hot and Energetic Universe: A White Paper presenting the science
  theme motivating the Athena+ mission

\bibitem[{{Nandra} {et~al.}(2007){Nandra}, {O'Neill}, {George}, \&
  {Reeves}}]{pexmon}
{Nandra}, K., {O'Neill}, P.~M., {George}, I.~M., \& {Reeves}, J.~N. 2007,
  \mnras, 382, 194

\bibitem[{{Nelson} {et~al.}(2004){Nelson}, {Green}, {Bower}, {Gebhardt}, \&
  {Weistrop}}]{nelson04}
{Nelson}, C.~H., {Green}, R.~F., {Bower}, G., {Gebhardt}, K., \& {Weistrop}, D.
  2004, \apj, 615, 652

\bibitem[{{Netzer}(1990)}]{netzer90_saasfee}
{Netzer}, H. 1990, in Active Galactic Nuclei, ed. R.~D. {Blandford},
  H.~{Netzer}, L.~{Woltjer}, T.~J.-L. {Courvoisier}, \& M.~{Mayor}, 57--160

\bibitem[{{Oliva} {et~al.}(1999){Oliva}, {Origlia}, {Maiolino}, \&
  {Moorwood}}]{oliva99}
{Oliva}, E., {Origlia}, L., {Maiolino}, R., \& {Moorwood}, A.~F.~M. 1999, \aap,
  350, 9

\bibitem[{{Onken} {et~al.}(2004){Onken}, {Ferrarese}, {Merritt}, {Peterson},
  {Pogge}, {Vestergaard}, \& {Wandel}}]{onken04}
{Onken}, C.~A., {Ferrarese}, L., {Merritt}, D., {Peterson}, B.~M., {Pogge},
  R.~W., {Vestergaard}, M., \& {Wandel}, A. 2004, \apj, 615, 645

\bibitem[{{Onken} {et~al.}(2014){Onken}, {Valluri}, {Brown}, {McGregor},
  {Peterson}, {Bentz}, {Ferrarese}, {Pogge}, {Vestergaard}, {Storchi-Bergmann},
  \& {Riffel}}]{onken14}
{Onken}, C.~A., {Valluri}, M., {Brown}, J.~S., {McGregor}, P.~J., {Peterson},
  B.~M., {Bentz}, M.~C., {Ferrarese}, L., {Pogge}, R.~W., {Vestergaard}, M.,
  {Storchi-Bergmann}, T., \& {Riffel}, R.~A. 2014, \apj, 791, 37

\bibitem[{{Palmeri} {et~al.}(2003){Palmeri}, {Mendoza}, {Kallman}, {Bautista},
  \& {Mel{\'e}ndez}}]{palmeri03}
{Palmeri}, P., {Mendoza}, C., {Kallman}, T.~R., {Bautista}, M.~A., \&
  {Mel{\'e}ndez}, M. 2003, \aap, 410, 359

\bibitem[{{Paltani} \& {T{\"u}rler}(2005)}]{paltani05}
{Paltani}, S., \& {T{\"u}rler}, M. 2005, \aap, 435, 811

\bibitem[{{Peterson} {et~al.}(2004){Peterson}, {Ferrarese}, {Gilbert}, {Kaspi},
  {Malkan}, {Maoz}, {Merritt}, {Netzer}, {Onken}, {Pogge}, {Vestergaard}, \&
  {Wandel}}]{peterson04}
{Peterson}, B.~M., {Ferrarese}, L., {Gilbert}, K.~M., {Kaspi}, S., {Malkan},
  M.~A., {Maoz}, D., {Merritt}, D., {Netzer}, H., {Onken}, C.~A., {Pogge},
  R.~W., {Vestergaard}, M., \& {Wandel}, A. 2004, \apj, 613, 682

\bibitem[{{Petrucci} {et~al.}(2002){Petrucci}, {Henri}, {Maraschi}, {Ferrando},
  {Matt}, {Mouchet}, {Perola}, {Collin}, {Dumont}, {Haardt}, \&
  {Koch-Miramond}}]{petrucci02}
{Petrucci}, P.~O., {Henri}, G., {Maraschi}, L., {Ferrando}, P., {Matt}, G.,
  {Mouchet}, M., {Perola}, C., {Collin}, S., {Dumont}, A.~M., {Haardt}, F., \&
  {Koch-Miramond}, L. 2002, \aap, 388, L5

\bibitem[{{Ponti} {et~al.}(2013){Ponti}, {Cappi}, {Costantini}, {Bianchi},
  {Kaastra}, {De Marco}, {Fender}, {Petrucci}, {Kriss}, {Steenbrugge}, {Arav},
  {Behar}, {Branduardi-Raymont}, {Dadina}, {Ebrero}, {Lubi{\'n}ski},
  {Mehdipour}, {Paltani}, {Pinto}, \& {Tombesi}}]{ponti13}
{Ponti}, G., {Cappi}, M., {Costantini}, E., {Bianchi}, S., {Kaastra}, J.~S.,
  {De Marco}, B., {Fender}, R.~P., {Petrucci}, P.-O., {Kriss}, G.~A.,
  {Steenbrugge}, K.~C., {Arav}, N., {Behar}, E., {Branduardi-Raymont}, G.,
  {Dadina}, M., {Ebrero}, J., {Lubi{\'n}ski}, P., {Mehdipour}, M., {Paltani},
  S., {Pinto}, C., \& {Tombesi}, F. 2013, \aap, 549, A72

\bibitem[{{Pozo Nu{\~n}ez} {et~al.}(2015){Pozo Nu{\~n}ez}, {Ramolla},
  {Westhues}, {Haas}, {Chini}, {Steenbrugge}, {Barr Dom{\'{\i}}nguez},
  {Kaderhandt}, {Hackstein}, {Kollatschny}, {Zetzl}, {Hodapp}, \&
  {Murphy}}]{pozonunez15}
{Pozo Nu{\~n}ez}, F., {Ramolla}, M., {Westhues}, C., {Haas}, M., {Chini}, R.,
  {Steenbrugge}, K., {Barr Dom{\'{\i}}nguez}, A., {Kaderhandt}, L.,
  {Hackstein}, M., {Kollatschny}, W., {Zetzl}, M., {Hodapp}, K.~W., \&
  {Murphy}, M. 2015, \aa\ in press; arXiv:1502.06771

\bibitem[{{Ricci} {et~al.}(2014){Ricci}, {Ueda}, {Ichikawa}, {Paltani},
  {Boissay}, {Gandhi}, {Stalevski}, \& {Awaki}}]{ricci14a}
{Ricci}, C., {Ueda}, Y., {Ichikawa}, K., {Paltani}, S., {Boissay}, R.,
  {Gandhi}, P., {Stalevski}, M., \& {Awaki}, H. 2014, \aap, 567, A142

\bibitem[{{Richards} {et~al.}(2006){Richards}, {Lacy}, {Storrie-Lombardi},
  {Hall}, {Gallagher}, {Hines}, {Fan}, {Papovich}, {Vanden Berk}, {Trammell},
  {Schneider}, {Vestergaard}, {York}, {Jester}, {Anderson}, {Budav{\'a}ri}, \&
  {Szalay}}]{richards06}
{Richards}, G.~T., {Lacy}, M., {Storrie-Lombardi}, L.~J., {Hall}, P.~B.,
  {Gallagher}, S.~C., {Hines}, D.~C., {Fan}, X., {Papovich}, C., {Vanden Berk},
  D.~E., {Trammell}, G.~B., {Schneider}, D.~P., {Vestergaard}, M., {York},
  D.~G., {Jester}, S., {Anderson}, S.~F., {Budav{\'a}ri}, T., \& {Szalay},
  A.~S. 2006, \apjs, 166, 470

\bibitem[{{Scott} {et~al.}(2004){Scott}, {Kriss}, {Brotherton}, {Green},
  {Hutchings}, {Shull}, \& {Zheng}}]{scott04}
{Scott}, J.~E., {Kriss}, G.~A., {Brotherton}, M., {Green}, R.~F., {Hutchings},
  J., {Shull}, J.~M., \& {Zheng}, W. 2004, \apj, 615, 135

\bibitem[{{Shu} {et~al.}(2012){Shu}, {Wang}, {Yaqoob}, {Jiang}, \&
  {Zhou}}]{shu12}
{Shu}, X.~W., {Wang}, J.~X., {Yaqoob}, T., {Jiang}, P., \& {Zhou}, Y.~Y. 2012,
  \apjl, 744, L21

\bibitem[{{Shu} {et~al.}(2010){Shu}, {Yaqoob}, \& {Wang}}]{shu10}
{Shu}, X.~W., {Yaqoob}, T., \& {Wang}, J.~X. 2010, \apjs, 187, 581

\bibitem[{{Shu} {et~al.}(2011){Shu}, {Yaqoob}, \& {Wang}}]{shu11}
---. 2011, \apj, 738, 147

\bibitem[{{Sitko} {et~al.}(1993){Sitko}, {Sitko}, {Siemiginowska}, \&
  {Szczerba}}]{sitko93}
{Sitko}, M.~L., {Sitko}, A.~K., {Siemiginowska}, A., \& {Szczerba}, R. 1993,
  \apj, 409, 139

\bibitem[{{Suganuma} {et~al.}(2006){Suganuma}, {Yoshii}, {Kobayashi},
  {Minezaki}, {Enya}, {Tomita}, {Aoki}, {Koshida}, \& {Peterson}}]{suganuma06}
{Suganuma}, M., {Yoshii}, Y., {Kobayashi}, Y., {Minezaki}, T., {Enya}, K.,
  {Tomita}, H., {Aoki}, T., {Koshida}, S., \& {Peterson}, B.~A. 2006, \apj,
  639, 46

\bibitem[{{Takahashi} {et~al.}(2014){Takahashi}, {Mitsuda}, {Kelley},
  {Aharonian}, {Akamatsu}, {Akimoto}, {Allen}, {Anabuki}, {Angelini}, {Arnaud},
  \& et~al.}]{astroh14}
{Takahashi}, T., {Mitsuda}, K., {Kelley}, R., {Aharonian}, F., {Akamatsu}, H.,
  {Akimoto}, F., {Allen}, S., {Anabuki}, N., {Angelini}, L., {Arnaud}, K., \&
  et~al. 2014, in Society of Photo-Optical Instrumentation Engineers (SPIE)
  Conference Series, Vol. 9144, Society of Photo-Optical Instrumentation
  Engineers (SPIE) Conference Series, 25

\bibitem[{{Toba} {et~al.}(2014){Toba}, {Oyabu}, {Matsuhara}, {Malkan},
  {Gandhi}, {Nakagawa}, {Isobe}, {Shirahata}, {Oi}, {Ohyama}, {Takita},
  {Yamauchi}, \& {Yano}}]{toba14}
{Toba}, Y., {Oyabu}, S., {Matsuhara}, H., {Malkan}, M.~A., {Gandhi}, P.,
  {Nakagawa}, T., {Isobe}, N., {Shirahata}, M., {Oi}, N., {Ohyama}, Y.,
  {Takita}, S., {Yamauchi}, C., \& {Yano}, K. 2014, \apj, 788, 45

\bibitem[{{Tristram} {et~al.}(2009){Tristram}, {Raban}, {Meisenheimer},
  {Jaffe}, {R{\"o}ttgering}, {Burtscher}, {Cotton}, {Graser}, {Henning},
  {Leinert}, {Lopez}, {Morel}, {Perrin}, \& {Wittkowski}}]{tristram09}
{Tristram}, K.~R.~W., {Raban}, D., {Meisenheimer}, K., {Jaffe}, W.,
  {R{\"o}ttgering}, H., {Burtscher}, L., {Cotton}, W.~D., {Graser}, U.,
  {Henning}, T., {Leinert}, C., {Lopez}, B., {Morel}, S., {Perrin}, G., \&
  {Wittkowski}, M. 2009, \aap, 502, 67

\bibitem[{{Ursini} {et~al.}(2015){Ursini}, {Marinucci}, {Matt}, {Bianchi},
  {Tortosa}, {Stern}, {Ar{\'e}valo}, {Ballantyne}, {Bauer}, {Fabian},
  {Harrison}, {Lohfink}, {Reynolds}, \& {Walton}}]{ursini15}
{Ursini}, F., {Marinucci}, A., {Matt}, G., {Bianchi}, S., {Tortosa}, A.,
  {Stern}, D., {Ar{\'e}valo}, P., {Ballantyne}, D.~R., {Bauer}, F.~E.,
  {Fabian}, A.~C., {Harrison}, F.~A., {Lohfink}, A.~M., {Reynolds}, C.~S., \&
  {Walton}, D.~J. 2015, ArXiv e-prints

\bibitem[{{Vasudevan} \& {Fabian}(2009)}]{vasudevanfabian09}
{Vasudevan}, R.~V., \& {Fabian}, A.~C. 2009, \mnras, 392, 1124

\bibitem[{{Vestergaard}(2002)}]{vestergaard02}
{Vestergaard}, M. 2002, \apj, 571, 733

\bibitem[{{Wang} {et~al.}(2004){Wang}, {Watarai}, \& {Mineshige}}]{wang04}
{Wang}, J.-M., {Watarai}, K.-Y., \& {Mineshige}, S. 2004, \apjl, 607, L107

\bibitem[{{Weigelt} {et~al.}(2012){Weigelt}, {Hofmann}, {Kishimoto},
  {H{\"o}nig}, {Schertl}, {Marconi}, {Millour}, {Petrov}, {Fraix-Burnet},
  {Malbet}, {Tristram}, \& {Vannier}}]{weigelt12}
{Weigelt}, G., {Hofmann}, K.-H., {Kishimoto}, M., {H{\"o}nig}, S., {Schertl},
  D., {Marconi}, A., {Millour}, F., {Petrov}, R., {Fraix-Burnet}, D., {Malbet},
  F., {Tristram}, K., \& {Vannier}, M. 2012, \aap, 541, L9

\bibitem[{{Woo} {et~al.}(2010){Woo}, {Treu}, {Barth}, {Wright}, {Walsh},
  {Bentz}, {Martini}, {Bennert}, {Canalizo}, {Filippenko}, {Gates}, {Greene},
  {Li}, {Malkan}, {Stern}, \& {Minezaki}}]{woo10}
{Woo}, J.-H., {Treu}, T., {Barth}, A.~J., {Wright}, S.~A., {Walsh}, J.~L.,
  {Bentz}, M.~C., {Martini}, P., {Bennert}, V.~N., {Canalizo}, G.,
  {Filippenko}, A.~V., {Gates}, E., {Greene}, J., {Li}, W., {Malkan}, M.~A.,
  {Stern}, D., \& {Minezaki}, T. 2010, \apj, 716, 269

\bibitem[{{Yaqoob}(2012)}]{yaqoob12}
{Yaqoob}, T. 2012, \mnras, 423, 3360

\bibitem[{{Yaqoob} {et~al.}(2003){Yaqoob}, {George}, {Kallman}, {Padmanabhan},
  {Weaver}, \& {Turner}}]{yaqoob03}
{Yaqoob}, T., {George}, I.~M., {Kallman}, T.~R., {Padmanabhan}, U., {Weaver},
  K.~A., \& {Turner}, T.~J. 2003, \apj, 596, 85

\bibitem[{{Yaqoob} \& {Padmanabhan}(2004)}]{yaqoob04}
{Yaqoob}, T., \& {Padmanabhan}, U. 2004, \apj, 604, 63

\bibitem[{{Zhou} {et~al.}(2010){Zhou}, {Zhang}, {Wang}, \& {Zhu}}]{zhou10}
{Zhou}, X.-L., {Zhang}, S.-N., {Wang}, D.-X., \& {Zhu}, L. 2010, \apj, 710, 16

\end{thebibliography}

\label{lastpage}
\end{document}